\documentclass[12pt]{article}
\usepackage{color,epsfig,multicol,graphicx,slashed}
\usepackage{amsmath}
\usepackage{hyperref}
\usepackage{amssymb}
\usepackage{amsfonts}
\usepackage{latexsym}
\newcommand{\ms}[2]{\,{\rm s}_{#2}(#1)\,}
\newcommand{\mc}[2]{\,{\rm c}_{#2}(#1)\,}
\newcommand{\mss}[2]{\,{\rm s}^2_{#2}(#1)\,}

\newcommand{\mcc}[2]{\,{\rm c}^2_{#2}(#1)\,}
\newcommand{\Dlr}{\stackrel{\leftrightarrow}{D}}

\newcommand{\Dslash}{\ensuremath{\slashed{D}}}
\newcommand{\pslash}{\ensuremath{\slashed{p}}}
\newcommand{\dslash}{\ensuremath{\slashed{\partial}}}
\newcommand{\Drsl}{\ensuremath{\stackrel{\rightarrow}{\slashed{D}}}}
\newcommand{\Aslash}{\ensuremath{\slashed{A}}}
\newcommand{\DD}{\Delta}
\newcommand{\cg}[1]{\,\mathcal{C}_{#1}\,}
\newcommand{\half}{{\textstyle \frac{1}{2}}}

\newcommand{\ggCF}{\frac{g^2\, C_F}{16 \pi^2} }

\newcommand{\psib}{\bar{\psi}}
\newcommand{\cgO}{\,\mathcal{C}_2\,}
\newcommand{\cgI}{\,\mathcal{C}_3\,}
\newcommand{\cgII}{\,\mathcal{C}_4\,}
\newcommand{\cgV}{\,\mathcal{C}_5\,}
\newcommand{\cgVIII}{\,\mathcal{C}_6\,}
\newcommand{\cgXIII}{\,\mathcal{C}_{1}\,}
\newcommand{\cgXVII}{\,\mathcal{C}_{7}\,}

\newcommand{\be}{\begin{equation}}
\newcommand{\ee}{\end{equation}}

\newcommand{\bea}{\begin{eqnarray}}
\newcommand{\eea}{\end{eqnarray}}
\newcommand{\MS}{{\overline{MS}}}

\newcommand{\itep}
{~\vspace{-1.2cm}

\begin{flushright}
{\normalsize DESY 08-034} \\
{\normalsize Edinburgh 2008/12} \\
{\normalsize Leipzig LU-ITP 2008/001} \\
{\normalsize Liverpool LTH 792}
\end{flushright}
\vspace{0.0cm}}

\begin{document}
\date{QCDSF Collaboration}
\begin{center}
\itep

{\Large\bf
Perturbative determination of $\mathbf{c_{SW}}$ for plaquette and\\[0.25em]
Symanzik gauge action and stout link clover fermions 
}
\vspace*{0.4cm}

{\large
R.~Horsley$^1$,
H.~Perlt$^{2}$,
P.~E.~L.~Rakow$^3$,
G.~Schierholz$^{4}$  and 
A.~Schiller$^2$
}

\vspace*{0.4cm}

{\sl
$^1$ School of Physics, University of Edinburgh, 
     Edinburgh EH9 3JZ, UK \\
$^2$ Institut f\"ur Theoretische Physik, Universit\"at
Leipzig, \\ D-04109 Leipzig, Germany \\
$^3$ Theoretical Physics Division, Department of Mathematical Sciences,
\\ University of Liverpool,  Liverpool L69 3BX, UK \\
$^4$ Deutsches Elektronen-Synchrotron DESY, 
     D-22603 Hamburg, Germany
}

\vspace*{0.4cm}

{\large QCDSF Collaboration}

\end{center}

\begin{abstract}
Using plaquette and Symanzik improved gauge action and stout link clover fermions
we determine the improvement coefficient $c_{SW}$  in one-loop lattice
perturbation theory from the off-shell quark-quark-gluon three-point function. 
In addition, we compute the  coefficients needed for the most general form
of quark field improvement and present the one-loop result for the critical 
hopping parameter $\kappa_c$. We discuss  mean field improvement for $c_{SW}$ 
and $\kappa_c$ and the choice of the mean field coupling for the actions we have considered.
\end{abstract}

\section{Introduction}

Simulations of Wilson-type fermions at realistic quark masses require an
improved action with good chiral properties and scaling behavior. 
A systematic improvement scheme that removes discretization errors order by
order in the lattice spacing $a$ has been proposed by
Symanzik~\cite{Symanzik:1983dc} and developed for on-shell quantities
in~\cite{Luscher:1984xn,Sheikholeslami:1985ij}. $\mathcal{O}(a)$ 
improvement of the Wilson fermion action is achieved by complementing
it with the so-called clover term~\cite{Sheikholeslami:1985ij}, provided the
associated clover coefficient is tuned properly.   

Wilson-type fermions break all chiral symmetries. This introduces an additive
negative mass renormalization term in the action, which gives rise to
singularities in the quark propagator at small quark masses and makes the
approach to the chiral regime difficult. A chiral improvement of the action is
expected to reduce the additive mass renormalization and the spread of
negative eigenvalues. Surprisingly, this is not accomplished by the clover
action. 

While the magnitude of the additive mass term decreases with
increasing clover term, the problem of negative eigenvalues is more severe for
the clover than for the standard Wilson action.
It is well known that via a combination of link fattening and tuning of the
clover coefficient, it is possible to reduce both the negative mass term and
the spread of negative
eigenvalues~\cite{DeGrand:1998mn,Boinepalli:2004fz,Capitani:2006ni}.  

The focus of this investigation is to determine the clover coefficient and the
additive mass renormalization for plaquette and Symanzik improved gauge action
and stout link clover fermions in one-loop lattice perturbation theory.

The Symanzik improved gauge action reads~\cite{Symanzik:1983dc}
\begin{equation}
  S_G^{\rm Sym} = \frac{6}{g^2} \,\,\left\{c_0
  \sum_{\rm Plaquette} \frac{1}{3}\, {\rm Re\, Tr\,}(1-U_{\rm Plaquette})
  \, +  c_1 \sum_{\rm Rectangle} \frac{1}{3}\,{\rm Re \, Tr\,}(1- U_{\rm
  Rectangle})\right\}
  \label{SG}
\end{equation}
with $c_0+8c_1=1$ and 
\begin{equation}
  c_0=\frac{5}{3}\,, \quad c_1=-\frac{1}{12}\,.
\end{equation}
This reduces to the standard plaquette action $S_G^{\rm Plaq}$ for $c_1=0$.

Clover fermions have the action for each quark flavor~\cite{Sheikholeslami:1985ij}
\begin{eqnarray} 
  S_F &=& a^4\, \sum_x \Big\{ - \frac{1}{2a} \, \left[\bar{\psi}(x)
  \widetilde U_\mu(x)\,(1-\gamma_\mu)\, \psi(x+a\hat{\mu})
  \right.
  \nonumber \\
  && 
  \hspace{8mm}\left.
  + \, \bar{\psi}(x) \widetilde U_\mu^\dagger(x-a\hat{\mu})\,(1+\gamma_\mu)\,
  \psi(x-a\hat{\mu})\right] 
  \label{SF}
  \\
  &&
  \hspace{8mm}
  + \, \frac{1}{a}\, (4 + a m_0 +a m)\, \bar{\psi}(x)\psi(x)
  - c_{SW}\, g\, \frac{a}{4}\, \bar{\psi}(x)\,
  \sigma_{\mu\nu} F_{\mu\nu}(x)\, \psi(x) \Big\} \,,
  \nonumber  
\end{eqnarray}
where
\begin{equation}
  am_0=\frac{1}{2\kappa_c} - 4 \,,
  \label{kc}
\end{equation}
$\kappa_c$ being the critical hopping parameter, is the additive mass
renormalization term, and $F_{\mu\nu}(x)$ is the field strength tensor in
clover form with
$\sigma_{\mu\nu}=(i/2)\,(\gamma_\mu\gamma_\nu-\gamma_\nu\gamma_\mu)$.  
We consider a version of clover fermions in which we do not smear links
in the clover term, but
the link variables $U_\mu$ in the next neighbor terms have been replaced by
(uniterated) stout links~\cite{Morningstar:2003gk}  
\begin{equation}
  \widetilde{U}_\mu(x) = e^{i\, Q_\mu(x)} \, U_\mu(x)
  \label{Ustout}
\end{equation}
with
\begin{equation}
  Q_\mu(x)=\frac{\omega}{2\,i} \left[V_\mu(x) U_\mu^\dagger(x) -
  U_\mu(x)V_\mu^\dagger(x) -\frac{1}{3} {\rm Tr} \,\left(V_\mu(x)
  U_\mu^\dagger(x) -  U_\mu(x)V_\mu^\dagger(x)\right)\right] \, .
\end{equation}
$V_\mu(x)$ denotes the sum over all staples associated with the link and
$\omega$ is a tunable weight factor. 
Stout smearing is preferred because (\ref{Ustout}) is expandable as 
a power series in $g^2$, so we can use perturbation theory. Many other
forms of smearing do not have this nice property. 
Because both the unit matrix and the $\gamma_\mu$ terms are smeared, 
each link is still a projection operator in the Dirac spin index. 

The reason for not smearing the clover term is that we want to keep 
the physical extent in lattice units of the fermion matrix small which is relevant 
for non-perturbative calculations.
In that respect we refer to these fermions as SLiNC fermions, from the phrase
{\bf S}tout {\bf Li}nk{\bf N}on-perturbative {\bf C}lover. 

The improvement coefficient $c_{SW}$ as well as the additive mass
renormalization $am_0$ are associated with the chiral limit. So we will carry
out the calculations for massless quarks, which simplifies things, though
it means that we cannot present values for the mass dependent corrections.

For complete $\mathcal{O}(a)$ improvement of the action there are five 
terms which would have to be added to the $\mathcal{O}(a)$ effective action, 
they are listed, for example, in \cite{Luscher:1996sc}. Fortunately, in 
the massless case only two remain, 
\begin{eqnarray}
  \mathcal{O}_1 &=& \psib \sigma_{\mu\nu} F_{\mu\nu} \psi\,, \\
  \mathcal{O}_2 &=& \psib \Dlr \Dlr \psi \,. 
\end{eqnarray}
The first is the clover term, the second is the Wilson mass term. 
We have both in our action, there is no need to add any other terms
to the action. 

In perturbation theory 
\begin{equation}
  c_{SW}=1 + g^2 \, c_{SW}^{(1)} + {\mathcal{O}(g^4)}\,.
  \label{csw}
\end{equation}
The one-loop coefficient $c_{SW}^{(1)}$ has been computed for the plaquette
action using twisted antiperiodic boundary conditions~\cite{Wohlert:1987rf}
and Schr\"odinger functional methods~\cite{Luscher:1996vw}. Moreover, using
conventional perturbation theory, Aoki and Kuramashi~\cite{Aoki:2003sj} have
computed $c_{SW}^{(1)}$ for certain improved gauge actions. All calculations
were performed for non-smeared links and limited to on-shell quantities.

We extend previous calculations of $c_{SW}^{(1)}$ to include stout links. This
is done by computing the one-loop correction to the off-shell quark-quark-gluon
three-point function. The improvement of the action is not sufficient to remove
discretization errors from Green functions. To achieve this, one must
also improve the quark fields. The most general form consistent with BRST 
symmetry is~\cite{Martinelli:2001ak}\footnote{In~\cite{Martinelli:2001ak}
the authors use $\Drsl$ and $\dslash$ instead of $\Drsl$ and $\Aslash$ - both
choices are equivalent.  Our choice is motivated by the discussion of
off-shell improvement in the next section.}
\begin{equation}
  \psi_{\star}(x)=\left(1 + a \,c_D \Drsl + a \,i\,g\,\,c_{NGI} \Aslash(x) \right) \,\psi(x)\,.
  \label{imppsi}
\end{equation}
{}From now we denote improved quark fields and
improved Green functions by an index~$\star$. These are made free of
$\mathcal O (a)$ effects by fixing the relevant improvement coefficients.

There is no {\it a priori} reason that the gauge variant contribution
$c_{NGI} \Aslash(x)$ vanishes. 
The perturbative expansion of $c_{NGI}$ has to start with the 
one-loop contribution~\cite{Martinelli:2001ak}.
As a byproduct of our calculation we determine that
coefficient $c_{NGI}^{(1)}$ 
\begin{equation}
  c_{NGI}=g^2\,c_{NGI}^{(1)} + {\mathcal{O}(g^4)}
  \label{cNGI}
\end{equation}
and find that it is indeed nonvanishing.

\section{Off-shell improvement}

It is known~\cite{Aoki:2003sj} that the one-loop contribution of the Sheikoleslami-Wohlert 
coefficient in conventional perturbation theory can be determined using the quark-quark-gluon 
vertex $\Lambda_\mu(p_1,p_2,c_{SW})$ sandwiched between {\sl on-shell} quark states.
$p_1$ ($p_2$) denotes the incoming (outgoing) quark momentum.
In general that vertex is an {\sl amputated} three-point Green function.

Let us  look at the ${\mathcal{O}}(a)$ expansion of tree-level 
$\Lambda^{(0)}_\mu(p_1,p_2,c_{SW})$ which is derived from action (\ref{SF})
\begin{equation}
  \Lambda^{(0)}_\mu(p_1,p_2,c_{SW}) = -i\, g  \,\gamma_\mu -g\, \half \, a\, 
  {\bf 1} (p_1 + p_2)_\mu 
  + c_{SW} \,i\, g\, \half \, a \,\sigma_{\mu \alpha} (p_1 -p_2)_\alpha\\
  +\mathcal{O}(a^2)\,.
  \label{treevertex}
\end{equation}
For simplicity we omit in all three-point Green functions the common overall 
color matrix $T^{a}$. That tree-level expression between on-shell quark states is free 
of order $\mathcal O (a)$ if the expansion of $c_{SW}$ starts with one,
as indicated in (\ref{csw}) 
\begin{equation}
  \bar u(p_2) \, \Lambda^{(0)}_{\star\mu}(p_1,p_2) \, u(p_1)  =  
  \bar u (p_2) \, (-i\,  g  \,\gamma_\mu )\, u(p_1)
  \,.
  \label{treeverteximproved}
\end{equation}
Therefore, at least  a one-loop calculation of the $\Lambda_\mu(p_1,p_2,c_{SW}^{(1)})$
is needed as necessary condition to determine $c_{SW}^{(1)}$. 

The {\sl off-shell} improvement condition states that the
{\sl non-amputated} improved quark-quark-gluon Green function $G_{\star \mu}(p_1,p_2,q)$ 
has to be free of $\mathcal{O}(a)$ terms in one-loop accuracy.
In position space that non-amputated improved quark-quark-gluon Green functions is defined 
via expectation values of improved quark fields and gauge fields as
\begin{equation}
  G_{\star\mu}(x,y,z)=\langle \psi_{\star}(x)\, \overline{\psi}_{\star}(y) \, A_\mu(z)\rangle
  \,.
\end{equation}
Since the gluon propagator is $\mathcal{O}(a)$-improved already, we do not need 
to improve gauge fields. Using relation (\ref{imppsi})
we can express the function  $G_{\star\mu}$ by the unimproved quark fields $\psi$ 
\begin{eqnarray}
  G_{\star\mu}(x,y,z)
  &=& G_{\mu}(x,y,z)+ a\,c_{D}\,\left\langle \left(\Dslash \Dslash^{-1}
  +\Dslash^{-1}\Dslash  \right)A_\mu \right\rangle
  \nonumber
  \\
  & & \quad\quad +\,  i \, a \, g \, c_{NGI}\,\left\langle\left(\Aslash \Dslash^{-1}
  +\Dslash^{-1}\Aslash  \right)A_\mu \right\rangle
  \,,
\end{eqnarray}
where $G_{\mu}(x,y,z)$ is the unimproved Green function.

Taking into account
\begin{equation}
  \left\langle \left(\Aslash \Dslash^{-1}+\Dslash^{-1}\Aslash  \right)A_\mu \right\rangle = 
  2\,a\,c_{D} \, \delta(x-y)\,\left\langle A_\mu(z) \right\rangle
\end{equation}
and setting $\langle A_\mu(z) \rangle=0$ (unless there is an unexpected symmetry breaking), 
we obtain the following relation between the improved and unimproved Green function
\begin{equation}
  G_{\star\mu}(x,y,z)= G_{\mu}(x,y,z)+i\,a \, g 
  \,c_{NGI}\,\left\langle\left(\Aslash \Dslash^{-1}+\Dslash^{-1}\Aslash  \right)
  A_\mu \right\rangle\,.
  \label{qqgimp2}
\end{equation}
{}From (\ref{qqgimp2}) it is obvious that tuning only $c_{SW}$ to its optimal value 
in $G_{\mu}(x,y,z)$, there would be an $\mathcal{O}(a)$ contribution
left in the improved Green function.
The requirement that $G_{\star\mu}(x,y,z)$ should be free of  ${\mathcal O}(a)$ terms
leads to an additional condition which determines the constant $c_{NGI}$. It has
not been calculated before.

Taking into account the expansion (\ref{cNGI}) of $c_{NGI}$ we get in momentum space 
($\mathcal{F}[\cdot]$ denotes the Fourier transform)
\begin{equation}
  i\, a \, g\, c_{NGI}\,\mathcal{F} \Big[\left\langle\left(\Aslash \Dslash^{-1} 
  +  \Dslash^{-1}\Aslash  \right)A_\mu 
  \right\rangle^{\rm tree}\Big] =
  i\, a \, g^3\, c_{NGI}^{(1)} \left(\gamma_\nu\frac{1}{i\, \pslash_1}
  +\frac{1}{i\, \pslash_2}\gamma_\nu\right)\, K^{\rm tree}_{\nu\mu}(q)\,,
  \label{cNGI1}
\end{equation}
or its amputated version
\begin{equation}
  i\, a \, g\, c_{NGI}\, \mathcal{F} \Big[\left\langle\left(\Aslash \Dslash^{-1} 
  +  \Dslash^{-1}\Aslash  \right)A_\mu 
  \right\rangle^{\rm tree}_{\rm amp}\Big] =
  -a \, g^3\, c_{NGI}^{(1)} \left(\pslash_2\, \gamma_\mu+\gamma_\mu\, \pslash_1\right)\,.
  \label{cNGI1amp}
\end{equation}

The relation between non-amputated and amputated unimproved and improved 
three-point Green functions are defined by
\begin{eqnarray}
  G_\mu(p_1,p_2,q)&=& S(p_2)\, \Lambda_\nu(p_1,p_2,q,c_{SW}^{(1)})\, S(p_1)\, K_{\nu\mu}(q)\,,
  \label{nonamp}
  \\
  G_{\star \mu}(p_1,p_2,q)&=& S_\star(p_2)\,  \Lambda_{\star\nu}(p_1,p_2,q) \, 
  S_\star(p_1) \, K_{\nu\mu}(q)
  \,,
  \label{nonampimp}
\end{eqnarray}
$K_{\nu\mu}(q)$ denotes the full gluon propagator which is $\mathcal{O}(a)$-improved 
already, $S(p)$ and $S_\star(p)$ the corresponding quark propagators.

With the definition of the quark self energy
\begin{equation}
  \Sigma(p)= \frac{1}{a} \Sigma_0 + i \, \pslash \, \Sigma_1(p) + \frac{a \, p^2}{2} \Sigma_2(p)
\end{equation}
the unimproved and improved inverse quark propagators are given by
\begin{eqnarray}
  S^{-1}(p)&=&i \, \pslash\, \Sigma_1(p) +\frac{a \,p^2}{2}\Sigma_2(p)=
        i \, \pslash \,\Sigma_1(p)\left(1-\frac{1}{2}a\, i \, \pslash\,
        \frac{\Sigma_2(p)}{\Sigma_1(p)}  \right)\,,
  \label{S}
  \\
  S_\star^{-1}(p)&=&i \, \pslash\, \Sigma_1(p)\,.
  \label{selfenergy}
\end{eqnarray}

Using the Fourier transformed  (\ref{qqgimp2}) with (\ref{cNGI1amp}) and amputating the Green 
function~(\ref{nonamp}), taking into account the inverse quark propagators (\ref{S}),
we get the off-shell improvement condition in momentum space
\begin{eqnarray}
  \Lambda_{\mu}(p_1,p_2,q,c_{SW}^{(1)})&=&\Lambda_{\star \mu}(p_1,p_2,q)+ 
        a \, g^3  c_{NGI}^{(1)} (\pslash_2 \, \gamma_\mu +\gamma_\mu\, \pslash_1)
  \nonumber\\
  & & \hspace{-0.7cm} -\, \frac{a}{2}\,i\, \pslash_2 \, \frac{\Sigma_2(p_2)}{\Sigma_1(p_2)}\, 
     \Lambda_{\star\mu}(p_1,p_2,q)
     -\frac{a}{2}\,\Lambda_{\star\mu}(p_1,p_2,q)\,  i\, \pslash_1 \, \frac{\Sigma_2(p_1)}{\Sigma_1(p_1)}
  \,.
\label{impcond}
\end{eqnarray}
This expression should hold to order $\mathcal{O}(g^3)$ by determining both 
$c_{NGI}^{(1)}$ and $c_{SW}^{(1)}$ correctly. It is clear from (\ref{impcond}) that the 
improvement term $\propto c_{NGI}^{(1)}$ does not contribute if both quarks are on-shell.

\section{The one-loop lattice quark-quark-gluon vertex}

\begin{figure}[!htb]
  \begin{center}
     \includegraphics[scale=0.3,width=0.8\textwidth]{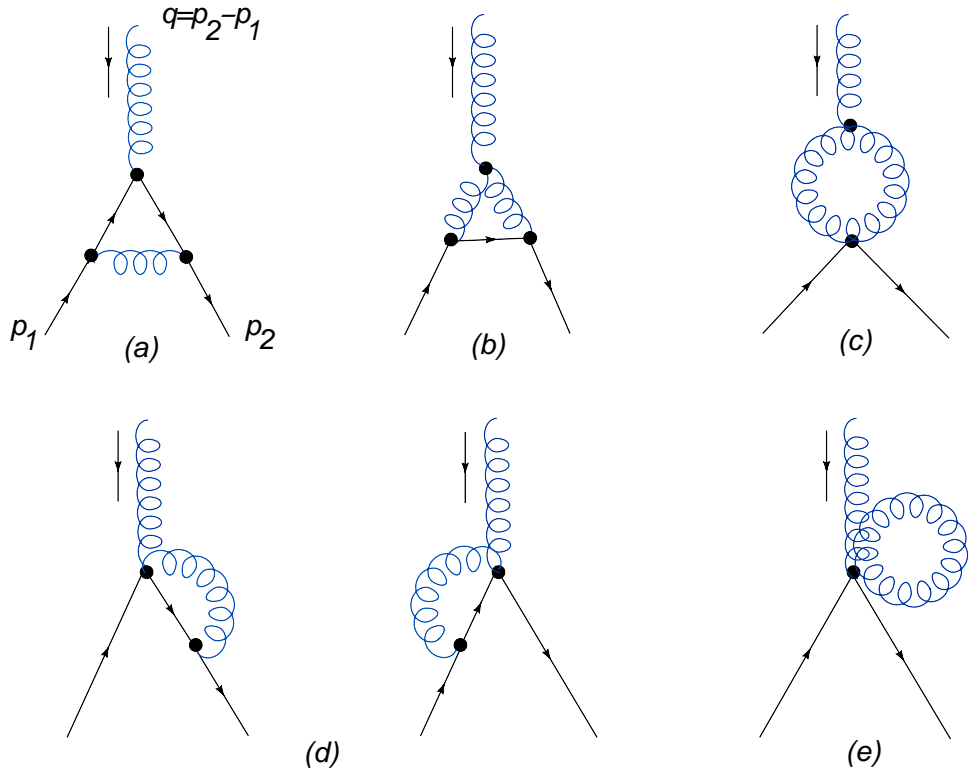}
  \end{center}
  \caption{One-loop diagrams contributing to the amputated quark-quark-gluon vertex.}
  \label{fig2}
\end{figure}

The diagrams contributing to the amputated one-loop three-point function are shown in Fig.~\ref{fig2}.
The calculation is performed with a mixture of symbolic and numerical techniques. For the symbolic
computation we use a {\it Mathematica} package that we developed for one-loop calculations
in lattice perturbation theory (for a more detailed description see ~\cite{Gockeler:1996hg}). 
It is based on an algorithm of Kawai et al.~\cite{Kawai:1980ja}. 
The symbolic treatment has several advantages: one can extract
the infrared singularities exactly and the results are given as functions of lattice integrals 
which can be determined with high precision. The disadvantage consists in very large expressions
especially for the problem under consideration. 
In the symbolic method the divergences are isolated by differentiation with respect to external momenta. 

Looking at the general analytic form of the gluon propagator for improved gauge actions~\cite{Horsley:2004mx}
one easily recognizes that a huge analytic expression would arise. As discussed in ~\cite{Horsley:2004mx}
we split the full gluon propagator $D_{\mu\nu}^{{\rm Sym}}(k,\xi)$ 
\begin{equation}
  D_{\mu\nu}^{{\rm Sym}}(k,\xi)=D_{\mu\nu}^{{\rm Plaq}}(k,\xi) + \Delta D_{\mu\nu}(k)\,,
  \label{Dprop}
\end{equation}
where $\xi$ is the covariant gauge parameter ($\xi=0$ corresponds to the Feynman gauge).
The diagrams with $D_{\mu\nu}^{{\rm Plaq}}(k,\xi)$ only contain the logarithmic parts and are treated
with our {\it Mathematica} package. The diagrams with at least one $\Delta D_{\mu\nu}(k)$
are infrared finite and can be determined safely with numerical methods. The decomposition 
(\ref{Dprop}) means that we always need to calculate the plaquette action result, as part of the calculation
for the improved gauge action. Therefore, we will give the results for both 
plaquette gauge action and Symanzik improved gauge action using the corresponding gluon propagators 
$D_{\mu\nu}^{{\rm Plaq}}$ and $D_{\mu\nu}^{{\rm Sym}}$, respectively.

Because the numerical part determines the accuracy of the total result we  discuss it in more detail. 
There are several possibilities to combine the various contributions of the one-loop diagrams as given 
in Fig.~\ref{fig2}. In view of a later analysis we have decided to group all coefficients in front 
of the independent color factors $C_F$ and $N_c$ and the powers of the stout parameter $\omega$
\begin{eqnarray}
  \Lambda^{{\rm num.}}_\mu &=& C_F\,\left(C_{0}+C_{1}\,\omega+C_{2}\,\omega^2+C_{3}\,\omega^3\right)+
  N_c\,\left(C_{4}+C_{5}\,\omega+C_{6}\,\omega^2+C_{7}\,\omega^3\right)\label{num2}\,,
\end{eqnarray}
where the $C_i$ have to be computed numerically. 
In order to obtain $C_i$ we first add all contributions of the diagrams shown in (\ref{fig2}) 
and integrate afterwards. We have used a Gauss-Legendre integration algorithm in four dimensions 
(for a description of the method see~\cite{Gockeler:1996hg}) and have chosen a sequence of small external
momenta $(p_1,p_2)$ to perform an extrapolation to vanishing momenta.

Let us illustrate this by an example:  the calculation of  the coefficient $C_4$.
We know the general structure of the one-loop amputated three-point function as (we set $a=1$)
\begin{eqnarray}
  M_\mu(p_1,p_2) &=& \gamma_\mu\, A(p_1,p_2) + {\rm\bf 1}\, p_{1,\mu}\, B(p_1,p_2) 
  + {\rm\bf 1}\, p_{2,\mu}\, C(p_1,p_2)  \nonumber\\
  & & + \,\sigma_{\mu\alpha}\,p_{1,\alpha} \,D(p_1,p_2) +
        \sigma_{\mu\alpha}\,p_{2,\alpha} \,E(p_1,p_2)\,.
\end{eqnarray}
{}From this we can extract the coefficients by the following projections
\begin{eqnarray}
  {\rm Tr}\,\gamma_\mu M_\mu &=& 4\,A(p_1,p_2), \quad\quad \mu \quad {\rm fixed}\,,
  \nonumber\\
  {\rm Tr}\,M_\mu &=&  4 \,p_{1,\mu}\, B(p_1,p_2) + 4 \,p_{2,\mu} \,C(p_1,p_2) \,,
  \\
  \sum_\mu\, {\rm Tr}\,\sigma_{\nu\mu}\,M_\mu &=& 12 \,p_{1,\nu} \,D(p_1,p_2) + 12 \,p_{2,\nu}\, E(p_1,p_2)\,.
  \nonumber
  \label{proj1}
\end{eqnarray}
Relations (\ref{proj1}) show that one has to compute the three-point function for all
four values of $\mu$. Further they suggest choosing the external momenta orthogonal
to each other: $p_1 \cdot p_2 = 0$. A simple choice is $p_{1,\mu}=(0,0,0,p_{1,4})$ and
$p_{2,\mu}=(0,0,p_{2,3},0)$. 

We discuss the determination of $B(p_1,p_2)$ and $C(p_1,p_2)$ in more detail.
For small momenta they can be described by the ansatz
\begin{eqnarray}
  B(p_1,p_2)&=& B_0 + B_1\,p_1^2 + B_2 \,p_2^2\,,
  \nonumber\\
  C(p_1,p_2)&=& C_0 + C_1\,p_1^2 + C_2 \,p_2^2\,.
  \label{BC}
\end{eqnarray}
The choice of the momenta is arbitrary except for two points. First, they should be
sufficiently small in order to justify ansatz (\ref{BC}). Second, they should not
be integer multiples of each other in order to avoid accidental symmetric results.
The symmetry of the problem demands the relation $B_0=C_0$ which must result from 
the numerical integration also. Performing the integration at fixed $p_1$ and $p_2$ we
obtain complex $4\times 4$ matrices for $M_3(p_1,p_2)$ and $M_4(p_1,p_2)$ from which
the quantities $B(p_1,p_2)$ and $C(p_1,p_2)$ are extracted via (\ref{proj1}).

\begin{figure}[!htb]
  \begin{center}
    \includegraphics[scale=0.01,width=0.8\textwidth]{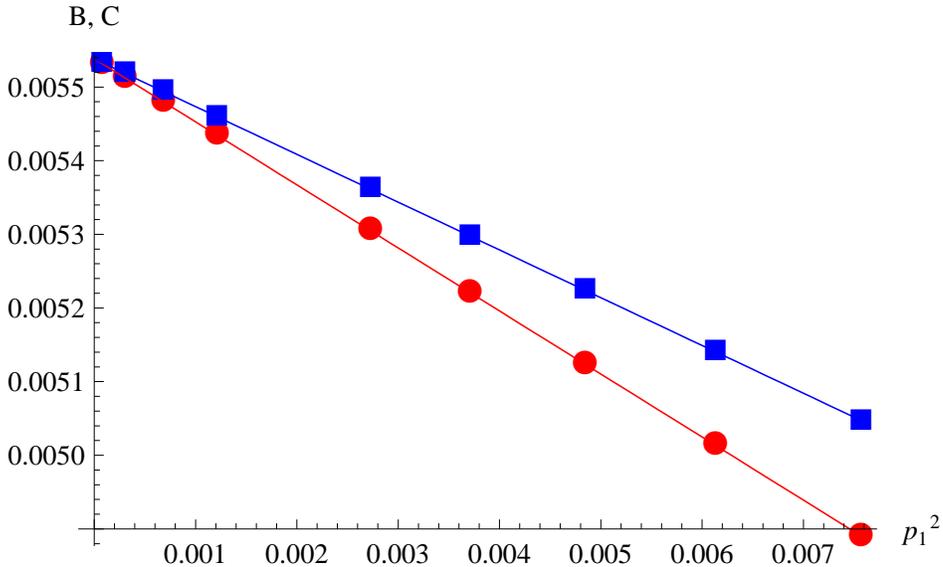}
  \end{center}
  \caption{$B(p_1,p_2)$ (circles) and $C(p_1,p_2)$ (squares) as function of $p_1^2$ together
           with their corresponding linear fits in $p_1^2$.}
  \label{fig3}
\end{figure}

A nonlinear regression fit with ansatz (\ref{BC}) gives
\begin{eqnarray}
  B_0&=&0.00553791 \quad {\rm with \,\, fit\,\, error}\,\, \delta B_0=7\times 10^{-8}\,,
  \nonumber\\
  C_0&=&0.00553789 \quad {\rm with \,\, fit\,\, error}\,\, \delta C_0=6\times 10^{-8}\,.
  \label{fitBC}
\end{eqnarray}
It shows that the symmetry is fulfilled up to an error of $\mathcal{O}(10^{-7})$
which sets one scale for the overall error of our numerical calculations. In Fig.~\ref{fig3} we
show the almost linear dependence of $B(p_1,p_2)$ and $C(p_1,p_2)$ on $p_1^2$.
(In the integration we haven chosen $p_{1,\mu}=0.87\, p_{2,\mu}$ so that we can restrict 
the plot to one variable.)

Another source of errors is the numerical Gauss-Legendre integration routine itself. 
We have chosen a sequence of $n^4=14^4$, $18^4$, $22^4$, $26^4$ and $30^4$ nodes in the 
four-dimensional hypercube and have performed an extrapolation to infinite nodes with an $1/n^4$ fit ansatz. 
Both procedures, Gauss-Legendre integration and the fit $p \rightarrow 0$, 
give a combined final error of $10^{-6}$.

The third error source are the errors of the lattice integrals of our {\it Mathematica} calculation 
for the terms containing the plaquette propagator $D_{\mu\nu}^{{\rm Plaq}}$ only.
These integrals have been calculated up to a precision of $\mathcal{O}(10^{-10})$. 
Therefore, their errors can be neglected in comparison with the others. 

Summarizing we find that the error of our numerical procedure is of $\mathcal{O}(10^{-6})$. 
Additionally, we have checked our results by an independent code which completely numerically computes 
the one-loop contributions for each diagram including the infrared logarithms. 
Both methods agree within errors.

\begin{figure}[!htb]
  \begin{center}
    \includegraphics[scale=0.01,width=0.8\textwidth]{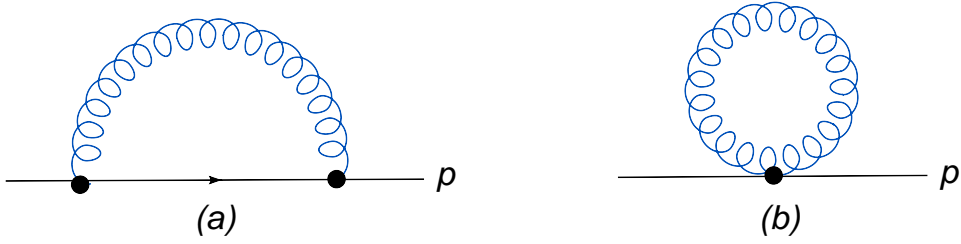}
  \end{center}
  \caption{One-loop diagrams contributing to the quark self energy.}
  \label{fig1}
\end{figure}

The Feynman rules for non-smeared Symanzik gauge action have been summarized in~\cite{Aoki:2003sj}.
For the stout smeared gauge links in the clover action the rules restricted to equal initial and 
final quark momenta are given in~\cite{Capitani:2006ni}.
As mentioned in the introduction we  perform a one-level smearing of the Wilson part
in the clover action. 
The corresponding Feynman rules needed for the one-loop quark-quark-gluon vertex are
much more complicated than those in~\cite{Capitani:2006ni}. The qqgg-vertex needed  
in diagrams (c) and (d) of Fig.~\ref{fig2} receives an additional antisymmetric piece.
The qqggg-vertex in diagram (e) does not even exist in the forward case. The Feynman rules are given
in Appendix A. The diagrams which are needed for the calculation of the quark propagator are
shown in Fig.~\ref{fig1}. We have performed our calculation in general covariant gauge.

\section{Results for the improvement coefficients and critical hopping parameter}

The anticipated general structure for the amputated three-point function in one-loop is
\begin{eqnarray}
  \Lambda_\mu(p_1,p_2,q)&=& \Lambda^{{\overline{MS}}}_\mu(p_1,p_2,q)
  +A_{\rm lat}\,i\,\frac{g^3}{16\pi^2}\,\gamma_\mu
  \nonumber\\
  & & + \, B_{\rm lat}\,\frac{a}{2}\,\frac{g^3}{16\pi^2}\,\left(\pslash_2\,\gamma_\mu
  +\gamma_\mu\,\pslash_1\right)
  + C_{\rm lat}\,i\,\frac{a}{2}\,\frac{g^3}{16\pi^2}\,\sigma_{\mu\alpha}\,q_\alpha \,.
  \label{Lam}
\end{eqnarray}
$\Lambda^{{\overline{MS}}}_\mu(p_1,p_2,q)$ is the universal part of the three-point function,
independent of the chosen gauge action, computed in the $\overline{MS}$-scheme
\begin{eqnarray}
  \Lambda^{{\overline{MS}}}_\mu(p_1,p_2,q)&=& -i\, g\, \gamma_\mu - 
  g\, \frac{a}{2}\,{\bf 1}\left( p_{1,\mu}+p_{2,\mu}\right)-
  c_{SW}\,i\, g\,\frac{a}{2}\sigma_{\mu\alpha}\,q_\alpha
  \nonumber\\
  & & + \,i\, \frac{1}{2}\,\frac{g^3}{16\pi^2}\,\Lambda^{{\overline{MS}}}_{1,\mu}(p_1,p_2,q) 
  + \frac{a}{2}\frac{g^3}{16\pi^2} \,\Lambda^{{\overline{MS}}}_{2,\mu}(p_1,p_2,q)\,.
  \label{LamMS}
\end{eqnarray}
We have calculated the complete expressions for $\Lambda^{{\overline{MS}}}_{1,\mu}(p_1,p_2,q)$ and $\Lambda^{{\overline{MS}}}_{2,\mu}(p_1,p_2,q)$.

The $\mathcal{O}(a)$ contribution, $\Lambda^{{\overline{MS}}}_{2, \mu}(p_1,p_2,q)$, simplifies
if we set $c_{SW}=1+\mathcal{O}(g^2)$ as in (\ref{csw}). After some algebra we find
\begin{eqnarray}
  \Lambda^{{\overline{MS}}}_{2,\mu}(p_1,p_2,q) &=& 
  \frac{1}{2}\left(\pslash_2\,\Lambda^{{\overline{MS}}}_{1,\mu}(p_1,p_2,q)+
  \Lambda^{{\overline{MS}}}_{1,\mu}(p_1,p_2,q)\,\pslash_1\right)
  \nonumber\\
  \label{lambda2}
  & & -\, C_F\left(\pslash_2\, \gamma_\mu \,(1-\xi)(1-\log(p_2^2/\mu^2))\right. 
  \\
  & & \quad \quad \, \left.  +\gamma_\mu\,\pslash_1 \,(1-\xi)(1-\log(p_1^2/\mu^2))\right)\,,
  \nonumber
\end{eqnarray}
where $\mu^2$ is the $\overline{MS}$ mass scale (not to be confused with the index $\mu$).
Therefore, we only need $\Lambda^{{\overline{MS}}}_{1,\mu}(p_1,p_2,q)$  to present 
the one-loop result (\ref{LamMS}). $\Lambda^{{\overline{MS}}}_{1,\mu}(p_1,p_2,q)$ is given in Appendix B.

If we insert (\ref{Lam}) and (\ref{LamMS}) with (\ref{lambda2}) into the off-shell
improvement relation (\ref{impcond}) we get the following conditions that all terms of order 
$\mathcal{O}(ag^3)$ have to vanish
\begin{eqnarray}
  \left(c_{SW}^{(1)} - \frac{C_{\rm lat}}{16\pi^2}\right)\,\sigma_{\mu\alpha}\,q_\alpha &=& 0\,,
  \label{cSWcond}\\
  \left(c_{NGI}^{(1)} - \frac{1}{32\pi^2}\,\left(A_{\rm lat}-B_{\rm lat}-\Sigma_{21}\right)\right)
        \left(\pslash_2 \, \gamma_\mu+ \gamma_\mu\, \pslash_1 \right) &=&0\,,
  \label{cNGIcond}
\end{eqnarray}
with $\Sigma_{21}$ defined from (\ref{S}) as
\begin{eqnarray}
  \frac{\Sigma_2(p)}{\Sigma_1(p)}&=&1+\frac{g^2\,C_F}{16\pi^2}\left((1-\xi)(1-\log(a^2p^2))+\Sigma_{21,0} \right)
  \nonumber\\
  &\equiv&1+\frac{g^2\,C_F}{16\pi^2}\left((1-\xi)(1-\log(p^2/\mu^2))\right)+\frac{g^2}{16\pi^2}\Sigma_{21}
  \label{SigmaWP}
\end{eqnarray}
and
\begin{equation}
  \Sigma_{21}=C_F\,\left( -(1-\xi)\log(a^2\mu^2)+\Sigma_{21,0} \right)\,.
  \label{SigmaWPr}
\end{equation}
The constant $\Sigma_{21,0}$ depends on the chosen lattice action.

It should be noted that  equations (\ref{cSWcond}) and (\ref{cNGIcond}) are obtained by using the general 
structure (\ref{lambda2}) only -- we do not need to insert the complete calculated result for $\Lambda^{{\overline{MS}}}_{1,\mu}(p_1,p_2,q)$.
In order to get momentum independent and gauge invariant improvement coefficients we see from (\ref{cSWcond}) that
$C_{\rm lat}$ itself has to be  constant and gauge invariant. From (\ref{cNGIcond}) and (\ref{SigmaWPr}) we
further conclude that the $\log(a^2\mu^2)$-terms from $A_{\rm lat}$ and $B_{\rm lat}$ have to cancel
those from $\Sigma_{21}$. The same is true for the corresponding gauge terms. The terms
$\propto(1-\xi)(1- \log(p_i^2/\mu^2))$ ($i=1,2$) coming from  (\ref{SigmaWP}) are canceled by the corresponding terms 
in (\ref{lambda2}).

Therefore, the relation between $\Lambda^{{\overline{MS}}}_{1,\mu}(p_1,p_2,q)$ and 
$\Lambda^{{\overline{MS}}}_{2,\mu}(p_1,p_2,q)$ as given in (\ref{lambda2}) is a nontrivial result.
Once more, it should be emphasized that this relation only holds if we use $c_{SW}=1$
at leading order in $g^2$. If (\ref{lambda2}) were not true, we would not be able to improve
the Green functions by adding the simple $\mathcal{O}(a)$ terms we have considered.

For completeness we also give the corresponding one-loop values for the quark field improvement
coefficient $c_D$ as defined in (\ref{imppsi}). They can be derived from the $\mathcal{O}(a)$ improvement 
of the quark propagator. The one-loop improvement coefficient $c_D^{(1)}$ is related to the quark self
energy by
\begin{equation}
  c_D = -\frac{1}{4}\,\left(1+\ggCF \, \left(2\,\Sigma_1-\Sigma_2\right)\right)
  +\mathcal{O}(g^4)\equiv  -\frac{1}{4}\,\left(1+g^2\,c_D^{(1)}\right)+\mathcal{O}(g^4)\,.
  \label{cD1}
\end{equation}
$c_D^{(1)}$ has been calculated for ordinary clover fermions and plaquette gauge
action in~\cite{Capitani:2000xi}.

Now we present our numerical results for general covariant gauge $\xi$ and as function of the 
stout parameter $\omega$. For the plaquette action with stout smearing the quantities
$A_{\rm lat}$, $B_{\rm lat}$ and $C_{\rm lat}$ are obtained as
\begin{eqnarray}
  A_{\rm lat}^{\rm Plaq}&=&C_F\,\Big(9.206269 +3.792010\,\xi - 196.44601\,\omega + 739.683641\,\omega^2 
  \nonumber \\
  &&
  \quad \quad + (1-\xi)\log (a^2\mu^2)   \Big)
  \nonumber\\
  & &+ \,
  N_c\,\left(-4.301720  + 0.693147\,\xi  + \,(1-\xi/4)\log (a^2\mu^2)   \right)\,,
  \nonumber\\
  B_{\rm lat}^{\rm Plaq}&=&C_F\,\Big(9.357942 + 5.727769\,\xi  - 208.583208\,\omega + 711.565256\,\omega^2 
  \nonumber \\
  &&
  \quad \quad  + 2\,(1-\xi)\log (a^2\mu^2)   \Big)
  \\
  & &+\,
  N_c\,\left(-4.752081 +0.693147\,\xi +3.683890\,\omega  + (1-\xi/4)\log (a^2\mu^2) \right)\,,
  \nonumber\\
  C_{\rm lat}^{\rm Plaq}&=&C_F\,\left(26.471857 + 170.412296\,\omega - 582.177099\,\omega^2\right)
  \nonumber\\ 
  & &+\,
  N_c\,\left(2.372649  + 1.518742\,\omega -44.971612\,\omega^2\right)\,.
  \nonumber
  \label{ABCplaq}
\end{eqnarray}
For the stout smeared Symanzik action we get
\begin{eqnarray}
  A_{\rm lat}^{\rm Sym}&=&C_F\,\Big(5.973656 +3.792010\,\xi - 147.890719\,\omega + 541.380348\,\omega^2 
  \nonumber \\
  && + \, 
  (1-\xi)\log (a^2\mu^2)   \Big)
  \nonumber\\
  & &+ \,
  N_c\,\left(-3.08478  + 0.693159\,\xi - 0.384236\,\omega  + (1-\xi/4)\log (a^2\mu^2)   \right)\,,
  \nonumber\\
  B_{\rm lat}^{\rm Sym}&=&C_F\,\Big(6.007320 + 5.727769\,\xi  - 163.833410\,\omega + 542.892478\,\omega^2
  \nonumber \\
  &&
  + \, 2\,(1-\xi)\log (a^2\mu^2)   \Big)
  \\
  & &+\,
  N_c\,\left(-13.841082 +0.693179\,\xi +3.039641\,\omega  + (1-\xi/4)\log (a^2\mu^2) \right)\,,
  \nonumber \\
  C_{\rm lat}^{\rm Sym}&=&C_F\,\left(18.347163 + 130.772885\,\omega - 387.690744\,\omega^2\right)
  \nonumber\\ 
  & &+\,
  N_c\,\left(2.175560 + 2.511657\,\omega -50.832203\,\omega^2\right)\,.
  \nonumber
  \label{ABC}
\end{eqnarray}
As shown in (\ref{impcond}) (or equivalently (\ref{cNGIcond})) we need 
the self energy parts $\Sigma_1(p)$ and  $\Sigma_2(p)$ as defined in (\ref{S}) to solve the
off-shell improvement condition. They have the general form
\begin{eqnarray}
  \Sigma_{1}(p)&=&1-\ggCF \, \left[(1-\xi)\log (a^2p^2) +\Sigma_{1,0}\right]\,,\nonumber\\
  \Sigma_{2}(p)&=&1-\ggCF \, \left[2\,(1-\xi)\log (a^2p^2) +\Sigma_{2,0}\right]\,.
  \label{SigmapW}
\end{eqnarray}
For the  plaquette and Symanzik actions we obtain
\begin{eqnarray}
  \Sigma_{1,0}^{\rm Plaq} &=& 8.206268 - 196.446005\,\omega + 739.683641\,\omega^2+4.792010\,\xi\,,\nonumber\\
  \Sigma_{2,0}^{\rm Plaq} &=& 7.357942 - 208.583208\,\omega + 711.565260\,\omega^2+7.727769\,\xi\,,\nonumber\\
  \Sigma_{1,0}^{\rm Sym}  &=& 4.973689 - 147.890720\,\omega + 541.380518\,\omega^2+4.792010\,\xi\,,\\
  \Sigma_{2,0}^{\rm Sym}  &=& 4.007613 - 163.833419\,\omega + 542.892535\,\omega^2+7.727769\,\xi\,.\nonumber
  \label{sigmas}
\end{eqnarray}
This results in the following expressions for $\Sigma_{21}$ as defined in (\ref{SigmaWPr})
\begin{eqnarray}
  \Sigma_{21}^{\rm Plaq} &=& C_F\, \Big(-0.151673 - 1.935759\,\xi + 12.137203\,\omega + 28.118384\,\omega^2
  \nonumber\\
  & & \quad\quad\quad -\,(1-\xi)\,\log(a^2\mu^2)\Big)\,,
  \nonumber\\
  \Sigma_{21}^{\rm Sym} &=& C_F\, \Big(-0.033924  - 1.935759\,\xi + 15.942699\,\omega-1.512017\,\omega^2 
  \\
  & & \quad\quad\quad -\, (1-\xi)\,\log(a^2\mu^2)\Big)\,.
  \nonumber
  \label{SigmaWPnum}
\end{eqnarray}

Inserting the corresponding numbers into (\ref{cSWcond}), (\ref{cNGIcond}) and (\ref{cD1}), 
we obtain the one-loop contributions of the clover improvement coefficient
\begin{eqnarray}
  c_{SW}^{(1),{\rm Plaq}}&=&C_F\,\left(0.167635 + 1.079148\,\omega - 3.697285\,\omega^2\right)
  \nonumber\\ 
  & &+\,
  N_c\,\left(0.015025 + 0.009617\,\omega - 0.284786\,\omega^2\right)\,,
  \label{cswplaq}\\
  c_{SW}^{(1),{\rm Sym}}&=&C_F\,\left(0.116185 + 0.828129\,\omega - 2.455080\,\omega^2\right)
  \nonumber\\ 
  & &+\,
  N_c\,\left(0.013777 + 0.015905\,\omega - 0.321899\,\omega^2\right)\,,
  \label{cswSym}
\end{eqnarray}
the off-shell quark field improvement coefficient
\begin{eqnarray}
  c_{NGI}^{(1),{\rm Plaq}}&=& N_c\,\left(0.001426  - 0.011664 \,\omega \right)\,,
  \label{cNGIplaq}\\
  c_{NGI}^{(1),{\rm Sym}}&=& N_c\,\left(0.002395 - 0.010841\,\omega \right)\,,
  \label{cNGISym}
\end{eqnarray}
and  the on-shell quark field improvement coefficient
\begin{eqnarray}
  c_D^{(1),{\rm Plaq}}&=& C_F\,\left( 0.057339 + 0.011755\,\xi -
   1.167149\,\omega + 4.862163\,\omega^2\right)\,,
  \label{cD2}\\
  c_D^{(1),{\rm Sym}} &=& C_F\,\left(0.037614 + 0.011755\,\xi - 0.835571\,\omega +
   3.418757\,\omega^2 \right)\,,
  \label{cD3}
\end{eqnarray}
for the plaquette and Symanzik action, respectively.
For $\omega=0$ both the plaquette result (\ref{cswplaq}) and the Symanzik result
(\ref{cswSym}) agree, within the accuracy of our calculations,  
with the numbers quoted in~\cite{Wohlert:1987rf,Luscher:1996vw} and~\cite{Aoki:2003sj}.

{}From Ward identity considerations it is known that the coefficient $c_{NGI}$ has to be 
proportional to $N_c$ only. Additionally, $c_{NGI}$ and $c_{SW}$ should be gauge invariant.
Both conditions are fulfilled within the errors which have been discussed in the previous section.
It should be noted that (\ref{cNGIplaq}) and (\ref{cNGISym}) are the first one-loop results for 
the quark field improvement coefficient $c_{NGI}$.
The gauge dependent improvement coefficient $c_D$ depends only on the color factor $C_F$
because it is determined by $\mathcal{O}(a)$ improvement of the quark propagator.

The additive mass renormalization is given by
\begin{equation}
  am_0=\ggCF \,\frac{\Sigma_0}{4} \,.
\end{equation}
This leads to the critical hopping parameter $\kappa_c$, at which chiral
symmetry is approximately restored, 
\begin{equation}
  \kappa_c=\frac{1}{8}\left( 1-
  \ggCF \,\frac{\Sigma_0}{4}\right)\,.
  \label{kappac}
\end{equation}
Using the plaquette or Symanzik gauge actions, we obtain
\begin{eqnarray}
  \Sigma_0^{\rm Plaq} &=& -31.986442 + 566.581765\,\omega -2235.407087\,\omega^2
  \,,
  \label{Sigma0plaq}
  \\ 
  \Sigma_0^{\rm Sym}  &=& -23.832351 + 418.212508\,\omega - 1685.597405\,\omega^2\,.
  \label{Sigma0sym}
\end{eqnarray}
This leads to the perturbative expression for $\kappa_c$
\begin{eqnarray}
  \kappa_c^{\rm Plaq} &=& \frac{1}{8} \left[
  1 + g^2 \, C_F \left(0.050639 - 0.896980 \,\omega + 3.697285 \,\omega^2
  \right) \right] 
  \,,
  \label{kappacplaq}
  \\
  \kappa_c^{\rm Sym} & =& \frac{1}{8}  \left[ 1   + g^2 \, C_F
  \left( 0.037730 - 0.662090\,\omega +2.668543\,\omega^2
  \right) \right]
  \,.
  \label{kappacSym}
\end{eqnarray}
For both actions $am_0$ can be tuned to zero for admissible values of $\omega$.
Using the smaller possible value we find $\omega=0.089396$ for the plaquette action 
and $\omega=0.088689$ for the Symanzik gauge action.

\section{Mean field improvement}

It is well known that one-loop perturbation theory in the bare coupling constant $g^2$ 
leads to a poor approximation. The coefficient of $g^2$ is large in most quantities,
and the series converges poorly. One traditional way to reduce this problem is 
by mean field improvement, which consists of two ideas.

The first is that we calculate each quantity in a simple mean field approximation,
and then re-express the perturbative result as the mean field result multiplied 
by a perturbative correction factor. If the mean field approximation is good,
the correction factor will be close to 1, and we have resolved the problem of the large
one-loop coefficient. As a good internal test of this part, we can simply look to see 
how large the coefficient in this correction factor is (the ``tadpole improved coefficient''),
compared with the initial unimproved coefficient. 

The second part of the mean field approximation is that we change our expansion parameter 
from the bare coupling $g^2$ to some ``boosted'' coupling constant, $g^2_{MF}$,
which we hope represents physics at a more relevant scale, and leads to a more rapidly convergent
series. A well-chosen boosted coupling would reduce the two-loop
coefficient. Unfortunately we usually cannot test this part of the improvement procedure, 
because the two-loop coefficient is unknown.  Fortunately, if the mean field approximation is good, 
the exact choice of boosted coupling constant will not be too crucial, because the lowest order 
improved coefficient will be a small number.

\subsection{Mean field approximation for smeared fermions} 

In the mean field approximation we typically assume that the gauge fields on each link are 
independently fluctuating variables, and that we can simply represent the links by an average
value $u_0$. Typical choices for $u_0$ would be to choose $u_0^4$ to be the average plaquette value, 
or to choose $u_0$ to be the average link value in the Landau gauge. 

A natural question is how we should extend the mean field approximation if we employ smearing. 
One possibility is to express everything in terms of two quantities, $u_0$, a mean value 
for the unsmeared link, and $u_S$, a mean value for smeared links\footnote{PR would like to thank 
Colin Morningstar for conversations on this point.}. 
We will discuss the relation between these two quantities later, first we want to make 
a general point about mean field approximations and smearing. 

The reason we smear our gauge links is to suppress very short range fluctuations in the gauge field, 
which is justified by the argument that these short range fluctuations are very 
lattice-dependent, rather than physical. However, put another way, suppressing short range fluctuations 
means that we are correlating nearby gauge links. So there is a certain tension between smearing 
and the mean field notion that each link is fluctuating independently. We will take the attitude that it
does still make sense to use the mean field approximation if smearing is mild -- but we should treat 
the results with some degree of caution if extreme smearing is used. 

Applying this double-$u$ mean field approximation to the SLiNC fermion matrix we find the following 
results for the principal fermion quantities, 
\begin{eqnarray} 
   && \Sigma_1(p) \approx u_S \,, \quad \Sigma_2(p) \approx u_S \,,
  \quad
   Z_\psi \approx u_S \,,
  \quad
  \kappa_c  \approx \frac{1}{8 u_S}\,, 
  \quad
   c_{SW}  \approx  \frac{u_S}{u_0^4}
\end{eqnarray}
(we define $Z_\psi$ by the relation $S^{\rm ren} = Z_\psi S^{\rm lat}$). 
For reasonable smearing we expect the smeared link $u_S$ to be  closer to 1 than the bare link $u_0$,
so most quantities will lie closer to their tree-level values with smearing.
However, the clover coefficient $c_{SW}$ is an exception; it will be further from 1 with smearing
than without, because we construct our clover term from unsmeared links. 

As a result, we obtain the mean field expressions for $\kappa_c$ and $c_{SW}$ by performing 
the following replacements
\begin{equation}
  \kappa_c(g^2) \rightarrow \kappa_c^{MF}(g_{MF}^2,u_S)=
        \frac{1}{8}\,\frac{u_S^{\rm pert}(g_{MF}^2)}{u_S}\,\kappa_c(g_{MF}^2)
\end{equation}
and
\begin{equation}
  c_{SW}(g^2) \rightarrow c_{SW}^{MF}(g_{MF}^2,u_S,u_0)=
        \frac{u_S}{u_0^4}\,\frac{u_0^{\rm pert}(g_{MF}^2)^{\,4}}{u_S^{\rm
        pert}(g_{MF}^2)}\,c_{SW}(g_{MF}^2)\,. 
\end{equation}
Here $u_S$ and $u_0$ are the measured smeared and unsmeared links at the given coupling and
$u_S^{\rm pert}$ and $u_0^{\rm pert}$ denote the corresponding expressions in lattice perturbation
theory.

\subsection{The smeared plaquette in perturbation theory} 

We will use $u_S^{\rm pert}$ derived from the smeared perturbative plaquette $P_S$ 
\begin{equation}
  u_S^{\rm pert} \equiv P_S^{1/4}. 
\end{equation} 
To one-loop order we have
\begin{equation}
  u_S^{\rm pert} = 1 - \ggCF\, k_S\,,
\end{equation}
with\footnote{We have written this integral for the case of a plaquette in the
1-2 plane, any orientation gives the same result.}
\begin{eqnarray}
  k_S = 8 \pi^2 a^4 \int \frac{d^4 k}{(2 \pi)^4}\hspace{-3mm} & D_{\alpha \beta}(k)  
  & \hspace{-3mm} \Bigl[\, V_{\alpha 1}(k,\omega) V_{\beta 1}(k,\omega) s_2^2(k) 
       +V_{\alpha 2}(k,\omega) V_{\beta 2}(k,\omega) s_1^2(k) 
  \nonumber \\ &  -& \hspace{-7mm} 
  \left(  V_{\alpha 1}(k,\omega) V_{\beta 2}(k,\omega)
  +  V_{\beta 1}(k,\omega) V_{\alpha 2}(k,\omega) \right)
  s_1(k) s_2(k) \Bigr] 
\end{eqnarray}
where $D_{\alpha \beta}(k)$ the gluon propagator for the action in question.
The smearing function $V_{\alpha \mu}(k,\omega)$ is defined in (\ref{Vdef}) in Appendix A, 
$s_\mu(k)$ and $s^2(k)$ used below are given in~(\ref{eq:A2}).
Using symmetry and the definition of $V$, the expression simplifies to 
\begin{equation} 
  k_S  = 16 \pi^2 a^4 \int \frac{d^4 k}{(2 \pi)^4} 
  \left[ D_{1 1}(k)  s_2(k) s_2(k) - D_{12}(k) s_1(k) s_2(k) \right]
  \left( 1 - 4 \, \omega \, s^2(k) \right)^2 
  \,.
  \label{SmearedPlaq}
\end{equation} 
We can see from this form that mild smearing has the effect of suppressing the contribution 
from large $k$. Setting $\omega = 0$ in $k_S$, we recover the unsmeared link in perturbation theory
\begin{equation}
  u_0^{\rm pert}= 1 - \ggCF\, k_S(\omega=0)\,.
  \label{u0}
\end{equation}

For the plaquette action propagator we can calculate the integral exactly. The result is
\begin{equation} 
  k_S^{\rm Plaq} = \pi^2 \left( 1 - 16 \,\omega + 72 \,\omega^2 \right) \,. 
\end{equation} 
Let us see how well this improves the expressions for $\kappa_c$ and $c_{SW}$.
Using the result ~(\ref{kappacplaq}) we find 
\begin{equation} 
  \kappa_c^{{\rm Plaq},MF} = \frac{1}{8 u_S}
  \left[ 1 + g_{MF}^2\, C_F \,
  \left( -0.011861 +  0.103020 \,\omega -0.802715\,\omega^2
 \right) \right]
\end{equation}
which successfully reduces the perturbative coefficients for every power of $\omega$.
Trying the same thing with the clover coefficient (\ref{cswplaq}) gives
\begin{eqnarray} 
  c_{SW}^{{\rm Plaq},MF} = \frac{u_S}{u_0^4}\, \Bigl\{ \hspace{-3mm} &1& \hspace{-3mm} + \, g^2_{MF}\, \Bigl[
   C_F\,\left(-0.019865 + 0.079148\,\omega +  0.813321\,\omega^2\right)
  \nonumber\\
  & &+\, 
  N_c\,\left(0.015025 + 0.009617\,\omega - 0.284786\,\omega^2\right)\,\Bigr]
  \Bigr\} \,. 
\end{eqnarray} 
Again, mean field improvement works well.

For the Symanzik action we calculate the integral in (\ref{SmearedPlaq}) numerically, and get the result
\begin{equation}
  k^{\rm Sym}_S = \pi^2 \left(  0.732525  -11.394696\,\omega 
  +   50.245225\,\omega^2 \right)\,.
\end{equation} 
The corresponding mean field improved  expressions for $\kappa_c$ (\ref{kappacSym}) 
and $c_{SW}$ (\ref{cswSym}) are
\begin{eqnarray} 
  \kappa_c^{{\rm Sym}, MF}   &=&  \frac{1}{8 u_S} 
  \left[1
  +  g_{MF}^2 \,C_F  \,   
  \left( -0.008053 +  0.0500781\,\omega -0.471784\,\omega^2 
  \right) \right]  
  \,,
  \\
  c_{SW}^{{\rm Sym},MF }& = & \frac{u_S}{u_0^4} 
  \Big\{ 1   +   g^2_{MF} \,
  \Big[
   C_F\,\left(-0.0211635 +  0.115961\,\omega +  0.685247\,\omega^2 \right)
  \nonumber\\
  & &+\, 
  N_c\,\left(0.013777 + 0.015905\,\omega - 0.321899\,\omega^2\right)\,\Big] 
  \Big\} \,. 
\end{eqnarray}

\subsection{Choice of $g^2_{MF}$} 

In this section we discuss the boosted coupling for $SU(3)$, we have set $N_c=3$, $C_F=4/3$ 
throughout.

From higher order continuum calculations we know that $g^2_{\MS}(\mu)$ is a good expansion parameter 
if  $\mu$ is close to the appropriate physical scale. On the other hand, series in the bare lattice 
coupling $g^2(a)$ usually converge poorly. To understand this difference let us compare the two 
couplings. To one-loop order we have
\begin{equation}
  \frac{1}{g^2_{\MS}(\mu)} - \frac{1}{g^2(a)} = 2b_0
  \left(\log\frac{\mu}{\Lambda_{\MS}} - \log\frac{1}{a\Lambda_{\rm lat}}\right) =
  2b_0 \log(a\mu) + d_g + N_f\, d_f \, ,
  \label{gg}
\end{equation}
where $b_0=(11-2N_f/3)/(4\pi)^2$, and $N_f$ is the number of flavors. 
The ratio of $\Lambda$ parameters is thus given by
\begin{equation}
  \frac{\Lambda_{\rm lat}}{\Lambda_{\MS}} = \exp \left(\frac{d_g + N_f\,
    d_f}{2b_0}\right) \, .
\end{equation}

The coefficient $d_g$ is known for the plaquette and Symanzik gauge action~\cite{Hasenfratz}:
\begin{equation}
  d_g^{\rm Plaq} = -0.4682\,, \quad d_g^{\rm Sym} = -0.2361 \,.
\end{equation}
In Appendix C we show that $d_f$ is independent of the stout smearing parameter $\omega$. 
Therefore, we can use the value for clover fermions computed in~\cite{Booth:2001qp}
\begin{equation}
  d_f=0.0314917 \,.
  \label{df}
\end{equation}
For $N_f=3$ this leads to
\begin{eqnarray}
  \frac{\Lambda_{\rm lat}}{\Lambda_{\MS}} &= 0.038 \quad
  \mbox{Plaquette}\,,\\ 
  \frac{\Lambda_{\rm lat}}{\Lambda_{\MS}} &= 0.289 \quad \mbox{Symanzik}\,.
\end{eqnarray}
These ratios are far from 1, especially for the plaquette action, which explains the poor 
convergence of series in $g^2(a)$.

Now let us see what happens to the Lambda ratio if we make the popular choice
of boosted coupling
\begin{equation}
  g_{MF}^2 = \frac{g^2}{u_0^4} \, .
  \label{gmf}
\end{equation}
Upon inserting (\ref{u0}) and (\ref{gmf}) in (\ref{gg}), we obtain
\begin{equation}
  \frac{1}{g^2_{\MS}(\mu)} - \frac{1}{g_{MF}^2(a)} = 2b_0
  \left(\log\frac{\mu}{\Lambda_{\MS}} - \log\frac{1}{a\Lambda_{\rm 
  lat}^{MF}}\right) = 2b_0 \log(a\mu) + d_g + N_f\, d_f +\frac{k_u}{3\pi^2} \, ,
\end{equation} 
which gives
\begin{equation}
  \frac{\Lambda_{\rm lat}^{MF}}{\Lambda_{\MS}} = \exp \left(\frac{d_g + N_f\,
    d_f + k_u/3\pi^2}{2b_0}\right) \, .
  \label{ratio}
\end{equation}
For $N_f=3$ the numerical values of this ratio are  
\begin{eqnarray}
  \frac{\Lambda_{\rm lat}^{MF}}{\Lambda_{\MS}} &= 0.702 \quad
  \mbox{Plaquette}\,,\\ 
  \frac{\Lambda_{\rm lat}^{MF}}{\Lambda_{\MS}} &= 2.459 \quad \mbox{Symanzik}\,.
\end{eqnarray}
We see that mean field improvement drives $\Lambda_{\rm lat}$ towards
$\Lambda_{\MS}$ for both the plaquette and Symanzik gauge action, giving
$g_{MF}^2 \approx g_{\MS}^2$, so that $g_{MF}^2$ appears to be a good
expansion parameter in both cases. A perfect match is obtained for $\mu=1/0.702
a$ ($\mu=1/2.459 a$) for the plaquette (Symanzik) action.

\section{Concluding remarks}

In the present paper we have computed the improvement coefficient $c_{SW}$ and the 
additive mass renormalization/critical hopping parameter in one-loop perturbation theory 
for general stout parameter $\omega$ performing a single smearing. 
To separate the effect of improving the gauge action from the effect of tuning 
the fermion action, we have done the calculation for both the plaquette action 
and the tree-level Symanzik gauge action. In addition we also present the $\mathcal{O}(g^2)$ 
corrections to the coefficients $c_{NGI}$ and $c_D$ needed to $\mathcal{O}(a)$ improve 
the quark fields in the most general case.

We give mean field (tadpole) improved results for $\kappa_c$ and $c_{SW}$. 
For both the plaquette and the Symanzik action the boosted coupling $g_{MF}^2$ turns out to be 
close to $g_{\MS}^2$, which makes $g_{MF}^2$ a good expansion parameter. 
We thus may expect that the perturbative series converges rapidly. 

For $N_f=3$ flavors of dynamical quarks it turns out that the one-loop
improved Symanzik gauge action~\cite{Luscher:1984xn} coincides largely with
its tree-level counterpart, with coefficients $c_0 \approx 5/3$, $c_1
\approx -1/12$ and $c_2 \approx 0$~\cite{Hao:2007iz}. This makes the
tree-level Symanzik action (\ref{SG}) stand out against other improved gauge
actions, at least from the perturbative point of view. 
SLiNC fermions represent a family of ultralocal, ultraviolet filtered clover fermions. 
While they share all prominent features of clover fermions, among them
$\mathcal{O}(a)$ improvement and flavor symmetry, they allow to further
optimize the chiral properties of the action by tuning the fattening of the
links. In our forthcoming simulations with $N_f=2+1$ and $2+1+1$ flavors of
dynamical quarks at realistic pion masses we shall employ this combination of
gauge and fermion actions.  

Knowing the perturbative (asymptotic) value of $c_{SW}$, we can derive a closed expression 
for $c_{SW}$ that covers the whole range of $g^2$. We will do so in a subsequent paper 
employing the Schr\"odinger functional method. The one-loop coefficient $c_{SW}^{(1)}$ 
varies only slightly within the interval $0 \leq \omega \leq 0.2$ for both the plaquette 
and Symanzik action. For $\omega=0.1$, which is our favorite value, the tadpole improved
one-loop coefficient becomes $c_{SW}^{(1)} \approx 0$, indicating that mean
field approximation works well. The final result is $c_{SW}^{MF} \approx u_S/u_0^4$ 
to a very good approximation for both gauge actions, where $u_S$ is the average 
smeared link, found by measuring the smeared plaquette, and $u_0$ the average unsmeared link,
found by measuring the unsmeared plaquette. 

This is to be compared with $c_{SW}^{MF} \approx 1/u_0^3$ over fermions with no smearing. 
We therefore expect $c_{SW}$ to be a steeper function of $g^2$ in the case of SLiNC fermions 
than for clover fermions.

Stout link fattening reduces the additive mass renormalization considerably,
with and without tadpole improvement, as expected. In fact, the critical
hopping parameter $\kappa_c$ can be tuned to its continuum value of $1/8$ for
an appropriate choice of $\omega$. We also confirm by early simulations with this 
action~\cite{preparation} that the spread of the
negative eigenvalues is reduced by a factor of $\approx 2$ for $\omega=0.1$
and non-perturbative $c_{SW}$, as compared to ordinary clover fermions.
SLiNC fermions have many other appealing features as well. 
The renormalization factors of quark bilinear operators, for example, come out to be very close
to unity, which hints at virtually continuum-like behavior.

\section*{Acknowledgment}

This investigation has been supported by DFG under contract FOR 465
(Forschergruppe Gitter-Hadronen-Ph\"anomenologie). We also acknowledge support
by the EU Integrated Infrastructure Initiative Hadron Physics (I3HP) under contract number
RII3-CT-2004-506078.

\renewcommand{\theequation}{A.\arabic{equation}}
\setcounter{equation}{0}

\section*{Appendix A: Feynman rules}

In this Appendix we give the Feynman rules for quark-gluon vertices derived from 
action (\ref{SF}) with single stout smeared gauge link variables in the Wilson part 
and general Wilson parameter $r$. The pieces in the vertices proportional to $c_{SW}$ 
are denoted with $\widetilde{V}$. They have been rederived  using our notations
and they agree with the Feynman rules given in~\cite{Aoki:2003sj}.
In the vertices we denote the incoming/outgoing quark momenta by $p_1/p_2$.
The incoming gluons are described by momenta $k_i$, Lorentz indices $\alpha,\beta,\gamma$ 
and color indices $a,b,c=1,\dots,N_c^2-1$. 

For the color matrices we have:
\begin{eqnarray}
  &&T^a T^b = \frac{1}{2 N_c} \delta^{ab} I_{N_c} + \frac{1}{2}( d^{abc}+ i \,f^{abc}) T^c
  \nonumber \\
  &&
   C_F = \frac{N_c^2-1}{2 N_c} \,, \quad 
  [T^a,T^b]=T^a T^b - T^b T^a\,, \quad \{T^a,T^b\}=T^a T^b + T^b T^a
  \\
  &&
  T_{ss}^{abc}=\{T^a,\{T^b,T^c\}\}
  \,, \quad
  T_{aa}^{abc}=[T^a,[T^b,T^c]]
  \,, \quad
  T_{sa}^{abc}=\{T^a,[T^b,T^c]\}
  \,.
  \nonumber
\end{eqnarray}
We use the abbreviations
\begin{eqnarray}
  &&\ms{k}{\mu}=\sin\left(\frac{a}{2}k_\mu\right), \quad \mc{k}{\mu}=\cos\left(\frac{a}{2}k_\mu\right)
  \,, \quad
  s^2(k) = \sum_\mu \mss{k}{\mu}
  \,, 
  \nonumber
  \\
  &&
  s^2(k_1,k_2)= \sum_\mu \ms{k_1+k_2}{\mu}\ms{k_1-k_2}{\mu} \equiv s^2(k_1)-s^2(k_2)\,.
  \label{eq:A2}
\end{eqnarray}
For later use we give the bare massless quark propagator
\begin{equation} 
  S(k) = \frac{a}{ i \sum_\mu \gamma_\mu \ms{2 k}{\mu}
  + r \sum_\mu \left( 1 - \mc{2k}{\mu} \right) }\,.
  \label{quarkprop}
\end{equation} 
The structure of the Wilson quark-gluon vertices is
\begin{eqnarray}
  W_{1\mu}(p_2,p_1)  &=& {i} \, \mc{p_2+p_1}{\mu} \,\gamma_\mu + r\,\ms{p_2+p_1}{\mu} \nonumber
  \\
  W_{2\mu}(p_2,p_1) &=&  {i}\, \ms{p_2+p_1}{\mu}\,\gamma_\mu - r\,\mc{p_2+p_1}{\mu}
  \label{eq:A3}
  \,.
\end{eqnarray}

Let us introduce the following functions to be useful in the definitions of the improved vertices
\begin{eqnarray}
  V_{\alpha\mu}(k,\omega)& =& \delta_{\alpha\mu} +  4\, \omega \, v_{\alpha\mu}(k) 
  \label{Vdef}\\
  v_{\alpha\mu}(k)&=&\ms{k}{\alpha}\ms{k}{\mu} -\delta_{\alpha\mu} \, s^2(k)
  \nonumber \\
  g_{\alpha\beta\mu}(k_1,k_2)&=&
  \delta_{\alpha\beta} \mc{k_1+k_2}{\alpha} \ms{k_1-k_2}{\mu}
  \nonumber\\
  &&-\,
  \delta_{\alpha\mu} \mc{k_2}{\alpha}\ms{2 k_1+k_2}{\beta}+
  \delta_{\beta\mu}  \mc{k_1}{\beta}\ms{2 k_2+k_1}{\alpha}
  \\
  w_{\alpha\mu}(k_1,k_2)&=& \ms{k_1+k_2}{\alpha}\ms{k_1-k_2}{\mu}- \delta_{\alpha\mu} \, s^2(k_1,k_2)\,,
  \\
  w_{\alpha\mu}(k,0)&=&v_{\alpha\mu}(k)\nonumber
\end{eqnarray}

\subsection*{The qqg-vertex: $V_\alpha^a(p_2,p_1,k_1; c_{SW},\omega)$}

The qqg-vertex including stout smeared links and clover contribution is given by the expression
($p_1+k_1=p_2$)
\begin{eqnarray}
  V_\alpha^a(p_2,p_1,k_1; c_{SW},\omega) &=& 
  - g\, T^a\, 
  \sum_\mu V_{\alpha\mu}(k_1,\omega)\, W_{1\mu}(p_2,p_1)+ c_{SW}\,\widetilde{V}_\alpha^a(k_1)\,.
\end{eqnarray}
The stout smeared part shows the separation property mentioned in~\cite{Capitani:2006ni}. 
The clover part is given by
\begin{eqnarray}
 \widetilde{V}_\alpha^a(k_1)&=& 
  -i\,g\, T^a\, \frac{r}{2}\,\sum_\mu \sigma_{\alpha\mu} \mc{k_1}{\alpha}\ms{2k_1}{\mu}\,.
\end{eqnarray}

\subsection*{The qqgg-vertex: $V_{\alpha\beta}^{ab}(p_2,p_1,k_1,k_2; c_{SW},\omega)$}

We define the qqgg-vertex as follows ($p_1+k_1+k_2=p_2$):
\begin{eqnarray}
  V_{\alpha\beta}^{ab}(p_2,p_1,k_1,k_2; c_{SW},\omega)=V_{\alpha\beta}^{\{a,b\}} + V_{\alpha\beta}^{[a,b]}+
	c_{SW}\,\widetilde{V}_{\alpha\beta}^{ab}(k_1,k_2)\,.
\end{eqnarray}
The stout smeared part is separated into two parts proportional to $\{T^a,T^b\}$ and $[T^a,T^b]$.
The anticommutator part shows the factorization property mentioned for two and four quark operators
\begin{eqnarray}
  V_{\alpha\beta}^{\{a,b\}} &=& \frac{1}{2} a\,g^2\,\{T^a,T^b\}\, 
  \sum_\mu V_{\alpha\mu}(k_1,\omega)\, V_{\beta\mu}(k_2,\omega)\,W_{2\mu}(p_2,p_1)
  \,.
\end{eqnarray}
The commutator part  is given by
\begin{eqnarray}
  V_{\alpha\beta}^{[a,b]}&=& \frac{1}{2} a\,g^2\,[T^a,T^b]\, 4 \, \omega
  \sum_\mu g_{\alpha\beta\mu}(k_1,k_2)\, \,W_{1\mu}(p_2,p_1) 
  \,.
  \label{eq:a12}
\end{eqnarray}
Note that this part is proportional to $\omega$. 
The part $\propto c_{SW}$ has been used in the form
\begin{eqnarray}
  \label{eq:a13}
 \widetilde{V}_{\alpha\beta}^{ab}(k_1,k_2)&=& i\,\frac{r}{4} a\,g^2\,[T^a,T^b]\,
  \Big\{2\, \sigma_{\alpha\beta}\big[2 \mc{k_1}{\beta}\mc{k_2}{\alpha}\mc{k_1+k_2}
  {\alpha}\mc{k_1+k_2}{\beta}
  \\
  && - \,
  \mc{k_1}{\alpha}\mc{k_2}{\beta}\big]+
  \delta_{\alpha\beta}\,\sum_\mu\,\sigma_{\alpha\mu}\ms{k_1+k_2}{\alpha}
  \left[\ms{2k_2}{\mu}-\ms{2k_1}{\mu}\right]\Big\}
  \,.
 \nonumber
\end{eqnarray}
Both (\ref{eq:a12}) and (\ref{eq:a13}) vanish for tadpole diagrams along quark lines.

\subsection*{The qqggg-vertex: $V_{\alpha\beta\gamma}^{abc}(p_2,p_1,k_1,k_2,k_3; c_{SW},\omega)$}

We present that vertex contribution in the following form ($p_1+k_1+k_2+k_3=p_2$)
\begin{eqnarray}
  &&\hspace{-20mm}V_{\alpha\beta\gamma}^{abc}(p_2,p_1,k_1,k_2,k_3; c_{SW},\omega)=\frac{1}{6} \,
  a^2 g^3 \times
  \nonumber\\
  &&
 \sum_\mu \bigg\{
  W_{1\mu}(p_2,p_1)\,\Big[F^{abc}_{\alpha\beta\gamma\mu}(k_1,k_2,k_3) + {\rm cyclic \ perm.}\Big]
  \nonumber \\
  && -\, 
  6 \,\omega \, W_{2\mu}(p_2,p_1) \, \Big[T_{sa}^{abc} \, 
  V_{\alpha\mu}(k_1) \, g_{\beta\gamma\mu}(k_2,k_3) + {\rm cyclic \ perm.} \Big] \bigg\} 
  \nonumber\\
  && +\,
   c_{SW}\, \widetilde{V}_{\alpha\beta\gamma}^{abc}(k_1,k_2,k_3)
  \,.
  \label{eq:A11}
\end{eqnarray}
Cyclic permutations have to be performed in  the gluon momenta as well as in the color 
and Lorentz indices of the three gluons. Note that the general stout smeared part is 
proportional both to $W_{1\mu}$ and $W_{2\mu}$.

The coefficient $F^{abc}_{\alpha\beta\gamma\mu}(k_1,k_2,k_3)$ is decomposed into its different 
color structures:
\begin{eqnarray}
  F^{abc}_{\alpha\beta\gamma\mu}(k_1,k_2,k_3)&=&
  T_{ss}^{abc} f^{(1)}_{\alpha\beta\gamma\mu}(k_1,k_2,k_3) + 
  T_{aa}^{abc}\, \big( f^{(2)}_{\alpha\beta\gamma\mu}(k_1,k_2,k_3)
  - f^{(2)}_{\alpha\gamma\beta\mu}(k_1,k_3,k_2)\big)
  \nonumber
  \\
  && +\,
  \left(T_{ss}^{abc}- \frac{1}{N_c} d^{abc}\right) f^{(3)}_{\alpha\beta\gamma\mu}(k_1,k_2,k_3)
  \,,
\end{eqnarray}
where the $f^{(i)}_{\alpha\beta\gamma\mu}$ are given as
\begin{eqnarray} 
  f^{(1)}_{\alpha\beta\gamma\mu}(k_1,k_2,k_3)&=&\frac{1}{2} \, 
  V_{\alpha\mu}(k_1,\omega)\, V_{\beta\mu}(k_2,\omega)\, V_{\gamma\mu}(k_3,\omega)
  \,,
  \nonumber
  \\
  f^{(2)}_{\alpha\beta\gamma\mu}(k_1,k_2,k_3)&=&
  \frac{1}{2} \, V_{\alpha\mu}(k_1,\omega)\, V_{\beta\mu}(k_2,\omega)\, \delta_{\gamma\mu}
  -\frac{1}{2} \,\delta_{\alpha\mu}\delta_{\beta\mu} \, V_{\gamma\mu}(k_3,\omega)
 \\
  &+&
  \hspace{-2mm}
  6 \, \omega \, \delta_{\alpha\beta} \Big[
  \mc{k_1-k_2}{\mu} \mc{2 k_3+k_1+k_2}{\beta}     \delta_{\gamma\mu}
  + \ms{k_3}{\mu} \ms{k_3 + 2 k_1}{\gamma}  \, \delta_{\beta\mu}
  \Big]
  \,,
  \nonumber
  \\
  f^{(3)}_{\alpha\beta\gamma\mu}(k_1,k_2,k_3)&=&
  2 \, \omega \, \delta_{\beta\gamma}
  \Big[\big( 3\, w_{\alpha\mu}(k_1,k_2+k_3) +  v_{\alpha\mu}(k_1+k_2+k_3)\big) \, \delta_{\alpha\beta}
  \nonumber
  \\
  &+&
  \hspace{-2mm}
  12 \ms{k_1}{\beta} \ms{k_2}{\alpha} \ms{k_3}{\alpha} 
  \big( \ms{k_1+k_2+k_3}{\beta} \, \delta_{\alpha\mu}- \ms{k_1+k_2+k_3}{\alpha} \delta_{\beta\mu}\big)
  \Big]
  \,.
  \nonumber
\end{eqnarray}

The clover part of the qqggg-vertex is given by
\begin{equation}
  \widetilde{V}_{\alpha\beta\gamma}^{abc}(k_1,k_2,k_3)=\frac{1}{6}\,
	\bigg\{\widetilde{\widetilde{V}}_{\alpha\beta\gamma}^{abc}(k_1,k_2,k_3) + {\rm total \ perm.}\bigg\}
  \label{Vtotclover}
\end{equation}
with
\begin{eqnarray}
  {\widetilde{\widetilde{V}}}_{\alpha\beta\gamma}^{abc}(k_1,k_2,k_3)&=&-3\,i\,g^3\,a^2\,r\times\nonumber\\
  && \hspace{-12mm}\bigg[T^aT^bT^c\delta_{\alpha\beta}\delta_{\alpha\gamma}
	\sum_\mu\,\sigma_{\alpha\mu}\bigg\{-\frac{1}{6}\mc{k_1+k_2+k_3}{\alpha}\ms{2(k_1+k_2+k_3)}{\mu}
  \nonumber\\
  && \hspace{-10mm} + \, \mc{k_1+k_2+k_3}{\alpha}\mc{k_1+k_2+k_3}{\mu}\mc{k_3-k_1}{\mu}\ms{k_2}{\mu}\bigg\}
  \nonumber\\
  && \hspace{-10mm} - \, \frac{1}{2}\bigg[T^aT^bT^c+T^cT^bT^a\bigg]\,\sigma_{\alpha\beta}\times
  \\
  && \hspace{-10mm} \bigg\{2\, \delta_{\beta\gamma}\mc{k_1+k_2+k_3}{\alpha}\mc{k_1+k_2+k_3}{\beta}
		\mc{k_3+k_2}{\alpha}\ms{k_1}{\beta} 
  \nonumber\\
  && \hspace{-8mm} + \, \delta_{\beta\gamma}\ms{k_3+k_2}{\beta}\mc{k_1+2k_2}{\alpha}
  \nonumber\\
  && \hspace{-8mm} + \, \delta_{\alpha\gamma}\ms{k_1+2k_2+k_3}{\alpha}\mc{k_1+k_2+k_3}
  {\beta}\mc{k_3-k_1}{\beta}\bigg\}\bigg]
  \nonumber\,.
\end{eqnarray}
In (\ref{Vtotclover}) the total permutation has to be performed in the gluon momenta, color and Lorentz indices.

We only need this vertex for the gluon tadpole diagram of Fig.~\ref{fig2}, which simplifies the expressions.
In the tadpole contribution to the vertex (\ref{eq:A11}) we denote the external gluon momentum  by
$q=p_2-p_1$, the  color index of the gluon by $a$ and the internal momenta by $k$ and $-k$.
The color indices ($b,c$) of the remaining gluons forming the tadpole are summed up using the color 
diagonality $\delta^{bc}$ of the gluon propagator, $k$ is the gluon momentum in the tadpole loop. 
So the stout smeared tadpole contribution is defined from the general qqggg-vertex 
(explicitly symmetrized in the three gluons) as
\begin{eqnarray}
  V_{\alpha\beta\gamma}^a(p_2,p_1,k)&=& \sum_{b=1}^{N_c^2-1}\bigg\{ 
  V_{\alpha\beta\gamma}^{a b b}(p_2,p_1,q,k, -k)+c_{SW}\,\widetilde{V}_{\alpha\beta\gamma}^{abb}(p_2,p_1,q,k,-k)\bigg\}
  \nonumber\\
  &=&\frac{1}{6} a^2\,g^3 \, T^a\,  \sum_\mu \, W_{1\mu}(p_2,p_1) V_{\alpha\beta\gamma\mu}(q,k)
  \\
  && +\,
  c_{SW}\,\sum_{b=1}^{N_c^2-1}\,\widetilde{V}_{\alpha\beta\gamma}^{abb}(p_2,p_1,q,k,-k)
  \,.\nonumber
\end{eqnarray}
Using that definition we obtain for the stout smeared part
\begin{eqnarray} 
  V_{\alpha\beta\gamma\mu}(q,k)&=&
  \Bigg\{ \left(6 \, C_F-N_c\right) \, f^{(1)}_{\alpha\beta\gamma\mu}(q,k,-k) + 
  \frac{N_c}{2}  
  \Big[
  f^{(2)}_{\beta\gamma\alpha\mu}(k,-k,q)-f^{(2)}_{\beta\alpha\gamma\mu}(k,q,-k)
  \nonumber\\
  && - \,
  f^{(2)}_{\gamma\alpha\beta\mu}(-k,q,k)+f^{(2)}_{\gamma\beta\alpha\mu}(-k,k,q)
  \Big]+
  4 \, C_F \, f^{(3)}_{\alpha\beta\gamma\mu}(q,k,-k)
  \\
  && + \,
  (4 \, C_F-N_c) \Big[ 
  f^{(3)}_{\beta\gamma\alpha\mu}(k,-k,q) +f^{(3)}_{\gamma\alpha\beta\mu}(-k,q,k)
  \Big]
  \Bigg\}
  \,.
  \nonumber
\end{eqnarray}

{}From that expression a convenient representation is found in the form
\begin{eqnarray}
  V_{\alpha\beta\gamma\mu}(q,k)& =& 
  \Bigg\{ \left(6 \, C_F-N_c\right) \,
  V_{\alpha\mu}(q,\omega)\, V_{\beta\mu}(k,\omega)\, V_{\gamma\mu}(k,\omega) 
  \nonumber\\
  && 
  \hspace{-3mm} +\,
  \frac{N_c}{2}  
  \Big[ 2 \,\delta_{\alpha\mu} V_{\beta\mu}(k,\omega)\, V_{\gamma\mu}(k,\omega) -
  V_{\alpha\mu}(q,\omega)\, \big( \delta_{\beta\mu} \, V_{\gamma\mu}(k,\omega) +
  \delta_{\gamma\mu} \, V_{\beta\mu}(k,\omega) \,
  \big) \Big] 
  \nonumber\\
  && 
  \hspace{-3mm}+\,
  2 \, \omega \, \Big[3  \left( 4 \, C_F-N_c \right) \, C_{\alpha\beta\gamma\mu}(q,k) + N_c
  \, D_{\alpha\beta\gamma\mu}(q,k) \Big]
  \Bigg\}
  \,.
\end{eqnarray}
The structures $C_{\alpha\beta\gamma\mu}$ and $D_{\alpha\beta\gamma\mu}$, 
additionally contributing to $O(\omega)$, are
\begin{eqnarray}
  C_{\alpha\beta\gamma\mu}(q,k)&=& - 4 
        \, \big[\delta_{\alpha\mu}   \mss{p}{\gamma}-\delta_{\alpha\gamma}\ms{p}{\alpha}\ms{p}{\mu}\big]\, 
           \big[\delta_{\beta \gamma}\mss{k}{\mu}   - \delta_{\beta \mu}   \ms{k}{\beta}\ms{k}{\gamma}\big] 
  \nonumber
  \\
  && - \,
  4 \, \delta_{\gamma\mu} \ms{p}{\beta}\ms{k}{\alpha}
  \big[ 
  \delta_{\alpha\beta} \ms{p}{\mu}\ms{k}{\mu } -\delta_{\alpha\mu} \ms{p}{\beta}\ms{k}{\beta} -
  \delta_{\beta\mu} \ms{p}{\alpha}\ms{k}{\alpha}
  \big]
  \nonumber
  \\
  && - \,
  \delta_{\alpha\mu}\delta_{\beta\mu}\delta_{\gamma\mu} \,
  \big[
  2 s^2(p)+ 2 s^2(k) - s^2(p+k) - s^2(p-k)
  \big] \,,
  \nonumber
  \\
  &&
  \\
  D_{\alpha\beta\gamma\mu}(q,k)&=& 
  - 3\, \delta_{\alpha\gamma}\delta_{\beta\mu} \mc{p+k}{\beta}\mc{p+k}{\gamma}   -
  3\, \delta_{\alpha\beta}\delta_{\gamma\mu} \mc{p-k}{\beta}\mc{p-k}{\gamma}
  \nonumber
  \\
  && + \,
  4 \,  \delta_{\beta\gamma}(\delta_{\alpha\beta}+\delta_{\beta\mu})  \ms{p}{\alpha}\ms{p}{\mu } +
  4 \,  \delta_{\alpha\mu}  (\delta_{\beta\mu}   +\delta_{\gamma\mu}) \ms{k}{\beta} \ms{k}{\gamma }
  \nonumber
  \\
  &&  - \,
  2\, \delta_{\alpha\mu} \delta_{\beta\mu} \delta_{\gamma\mu} \big[s^2(p)+s^2(k)\big] 
  +  6 \, \delta_{\alpha\mu}\delta_{\beta\gamma} \, \big[2 \mcc{p}{\gamma}\mcc{k}{\alpha} -1 \big] 
  \,.
  \nonumber
\end{eqnarray}

\renewcommand{\theequation}{B.\arabic{equation}}
\setcounter{equation}{0}

\section*{Appendix B: Three-point function - universal part}

As discussed above, the universal part of the three-point function has the form (\ref{lambda2})
when $c_{SW}=1 + \mathcal{O}(g^2)$.
Therefore, it is sufficient to give only the one-loop result for 
$\Lambda^{{\overline{MS}}}_{1,\mu}(p_1,p_2,q)$. It is cast into the following form ($q=p_2 - p_1$)
\begin{eqnarray}
  \Lambda^{{\overline{MS}}}_{1,\mu}(p_1,p_2,q) &=&F_1(p_1,p_2)\,\gamma_\mu+
  F_2(p_1,p_2)\,\pslash_2\, \gamma_\mu\pslash_1
  \nonumber
  \\& & +\,
   [ F_3(p_1,p_2) \,p_{1,\mu}+F_4(p_1,p_2) \,p_{2,\mu}]\, \pslash_1 
  \label{Blam1}
  \\& &
  +\,   [ F_5(p_1,p_2)\,p_{2,\mu}+F_6(p_1,p_2)\,p_{1,\mu}]\, \pslash_2 \,.
  \nonumber
\end{eqnarray}
Due to the symmetries $F_5(p_1,p_2)=F_3(p_2,p_1)$ and $F_6(p_1,p_2)=F_4(p_2,p_1)$ we have 
four independent functions $F_i(p_1,p_2)$ only. We represent them as follows:
\begin{eqnarray}
  F_1(p_1,p_2)&=&4\,C_F\,\xi - \frac{N_c}{2} (12+2\xi-\xi^2)
  + 2\,\Theta \left(\cgXIII\,\mathcal{S}+ N_c\,p_1.p_2 +  C_F \,q^2\right)
  \nonumber\\
  & &+\left(C_F(1-\xi)+\frac{N_c}{4}(4-\xi)\right)  \log\left( \frac{p_1^2p_2^2 }{\left(\mu^2\right)^2} \right)
  \\
  & &+\, V_1(p_1,p_2) \log \left( \frac{p_1^2}{q^2}\right)+V_1(p_2,p_1) \log\left( \frac{p_2^2}{q^2}\right)
  \,,
  \nonumber
\end{eqnarray}
\begin{eqnarray}
  \hspace{-1mm}F_2(p_1,p_2)=\frac{\Theta}{8}\left(2N_c(6-\xi)+ \cgO \frac{p_1.p_2\, q^2}{\DD}  \right)
  +\frac{\cgO}{4 \DD} \left[  
  p_1.q \log\left( \frac{p_1^2}{q^2}\right) - p_2.q   \log\left( \frac{p_2^2}{q^2}\right)
  \right] \,,
\end{eqnarray}
\begin{eqnarray}
  F_3(p_1,p_2)&=&\cgI\frac{p_2^2}{2\,\DD}+\frac{2\,N_c\,\xi}{q^2}
  +\frac{\Theta}{8\,\DD}
  \Big[ 
  4 N_c\,\xi (p_1.p_2)^2 
  +  \left(2 \cgI (6 \, \mathcal{S} + p_2^2)-\cgII  p_1.q      \right)\, p_2^2
  \Big]
  \nonumber\\
  & & 
  +\, \frac{1}{q^2}\left[V_2(p_1,p_2) \log\left( \frac{p_1^2}{q^2}\right)+V_3(p_1,p_2) \log\left( \frac{p_2^2}{q^2}\right)
  \right]
  \,,
\end{eqnarray}
\begin{eqnarray}
  F_4(p_1,p_2)&=&-\cgI \frac{p_1.p_2}{2\,\DD}-\frac{2\,N_c\,\xi}{q^2}
  + \frac{\Theta}{8\,\DD}
  \Big[ 
  4 (8\,C_F-  N_c (4-\xi))\, (p_1.p_2)^2 
  \nonumber\\ & &
  -\, (12 \cgI \mathcal{S} + 4 \cgVIII \, p_1^2 +(\cgV+ 8\, C_F(2+\xi)      )
  \, p_2^2) \, p_1.p_2
  +\cgXVII \, p_1^2\, p_2^2
  \Big]
  \\
  & &
  +\, \frac{1}{q^2}\left[
  V_4(p_1,p_2) \log\left( \frac{p_1^2}{q^2}\right)+V_5(p_1,p_2) \log\left( \frac{p_2^2}{q^2}\right)
  \right]
  \,.
  \nonumber
\end{eqnarray}
The function $V_i$ in front of the logarithms are found as follows
\begin{eqnarray}
  V_1(p_1,p_2) &=& C_F (3+\xi)-\frac{N_c}{4} (4-\xi)
  +\cgXIII \frac{ p_2.q \,p_1^2 }{\DD}  \nonumber\,,
  \nonumber
  \\
  V_2(p_1,p_2) &=& 
  \frac{1}{4 \, \DD}
  \Big[
  (4 \cgI-\cgII- 4 N_c \, \xi)\, p_2^2 \, q^2
  \nonumber \\ &&
  + \,
  (12 \cgI \mathcal{S}+ 4 N_c \, \xi \, p_1.p_2 + (\cgV + 8 \,C_F)\, q^2) \,p_2.q
  \Big]
  \,,
  \nonumber\\
  V_3(p_1,p_2) &=& 
  \frac{1}{4\,\DD\,p_1^2}
  \left[ -4 N_c\,\xi \, p_1.p_2 \, p_2.q \, p_1^2+\left( 
  -12 \cgI\mathcal{S} \, p_1.q +\cgII p_1^2 \, q^2
  \right) \,p_2^2
  \right]\,,
  \\
  V_4(p_1,p_2) &=& 
  V_2(p_2,p_1) + \frac{1}{4\,\DD} \left[ 
  - 8 \, C_F (1+\xi)\, p_1.q + ( 4 \, C_F (1- 3 \xi) + N_c (5-\xi) \xi)\,p_1^2
  \right]
  \,,
  \nonumber\\
  V_5(p_1,p_2) &=& V_2(p_1,p_2)
  +\frac{1}{4\,\DD}\left[
  (8\,C_F +N_c (2-\xi)\xi)\, p_2.q + (1+\xi) (4 \, C_F + N_c \, \xi) p_2^2
  \right]
  \,.\nonumber
\end{eqnarray}
We have introduced the kinematic functions
\begin{eqnarray}
  \DD &=&  (p_1.p_2)^2 - p_1^2\,p_2^2\,, \quad
  \mathcal{S} = \frac{p_1^2\,p_2^2\,q^2}{4\,\DD}\,,
  \nonumber \\
  \Theta &=& \frac{4}{\pi^2\,\sqrt{\DD}}\Bigg({\rm Sp}\left(\frac{p_2.q+\sqrt{\DD}}{p_2^2} \right)-
        {\rm Sp}\left(\frac{p_2.q-\sqrt{\DD}}{p_2^2} \right)
  \\
  &&
  + \, \frac{1}{2}\log\left( \frac{p_1.p_2-\sqrt{\DD}}{p_1.p_2
  +\sqrt{\DD}} \right) \log\left( \frac{q^2}{p_2^2} \right)\Bigg)\,,
  \nonumber
\end{eqnarray}
with ${\rm Sp}(x)$ being the Spence function:
$$
{\rm Sp}(x)=-\int_0^x\,dy \frac{\log (1-y)}{y}\nonumber\,.
$$
The quantities $\cg{i}$ depend on the color factors and gauge parameter and have the
values
\begin{eqnarray}
  \cgXIII &=& C_F\,(3+\xi)-\frac{1}{2}  N_c\,(1-\xi)\,,
  \nonumber \\
  \cgO &=& 8\,C_F + N_c\, (2+(3-\xi)\xi))\,,
  \nonumber\\
  \cgI &=& 4\,C_F\,(1+\xi)-N_c\,(4+(1-\xi)\xi)\,,
  \nonumber\\
  \cgII &=& 8\,C_F\,(2+\xi)-N_c\,(12+(4-3\xi)\xi)\,,\\
  \cgV &=& -N_c\,(4-(2+\xi)\xi)\,,
  \nonumber\\
  \cgVIII &=& 4\,C_F-N_c\,(1-\xi)\,,
  \nonumber\\
  \cgXVII&=& 8 \, C_F - N_c\, (16 -\xi^2)\,.
  \nonumber
\end{eqnarray}

In order to express the one-loop result (\ref{Blam1}) in terms of Spence functions, logarithms and 
rational functions of external momenta we have proceeded in two steps. First we have 
expanded all tensor integrals over the internal momentum into scalar three-point integrals times
tensor functions of the external momenta~\cite{Kizilersu:1995iz}. Then we used 
recursion relations of Davydychev~\cite{Davydychev:1992xr} to reduce these scalar three-point integrals into 
scalar two-point integrals and $\Theta$.

\renewcommand{\theequation}{C.\arabic{equation}}
\setcounter{equation}{0}

\section*{Appendix C: $\omega$-independence of $d_f$}

\begin{figure}[!htb]
  \begin{center}
    \includegraphics[scale=0.01,width=0.8\textwidth]{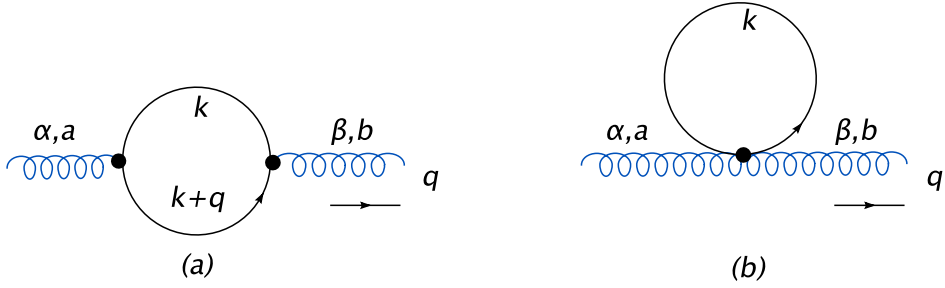}
  \end{center}
  \caption{One-loop quark vacuum polarization diagrams.}
  \label{fig4}
\end{figure}

We find $d_f$, the coefficient which tells us the fermionic shift in $\Lambda_{\rm lat}$, 
by calculating the massless quark vacuum polarization in a gluon with $a^2 q^2 \ll 1$:
\begin{eqnarray} 
  \lefteqn{ \Pi_{\alpha \beta}^{a b}(q; c_{SW}, \omega) =} \nonumber \\
  & -&  N_f \int \frac{d^4 k}{(2 \pi)^4}
  {\rm Tr} \left[ V_\alpha^a(q+k,k,q; c_{SW}, \omega) S(k)
  V_\beta^b (k, q+k, -q; c_{SW}, \omega) S(k+q) \right] \nonumber \\
  & -& N_f \int \frac{d^4 k}{(2 \pi)^4}
  {\rm Tr} \left[ V_{\alpha \beta}^{\{a,b\}}(k,k,q,-q; c_{SW}, \omega)
  S(k) \right]
  \,.
\end{eqnarray}  
The quark propagator $S$ and the vertices $V$ are defined in Appendix A, the trace here is over both spin and color. 
The corresponding one-loop diagrams are shown in Fig. \ref{fig4}.

In the required limit of small $a^2 q^2$ we can expand in  $q^2$ and drop any terms 
$\mathcal{O}(a^2q^4)$. We then get
\begin{eqnarray} 
  \Pi_{\alpha \beta}^{a b}(q; c_{SW}, \omega) &=& 
  \Pi_{\alpha \beta}^{a b}(q; c_{SW}, 0) -  2\, \omega\, N_f\,\delta^{a b} g^2 a^2 \times
  \nonumber\\
  & & \Bigg\{
  \sum_\mu\,\left(q_\alpha q_\mu - q^2 \delta_{\alpha \mu}\right)
  \int  \frac{d^4 k}{(2 \pi)^4} {\rm Tr} \left[
  W_{1 \mu}(k,k) S(k) W_{1 \beta}(k,k) S(k) \right] \nonumber \\
  & & +\, a
  \left(q_\alpha q_\beta - q^2 \delta_{\alpha \beta}\right)
  \int  \frac{d^4 k}{(2 \pi)^4} {\rm Tr} \left[
  W_{2 \beta}(k,k) S(k) \right]  \label{Pilong} \\
  & &+
  \sum_\mu\,\left(q_\beta q_\mu - q^2 \delta_{\beta \mu}\right)
  \int  \frac{d^4 k}{(2 \pi)^4} {\rm Tr} \left[
  W_{1 \alpha}(k,k) S(k) W_{1 \mu }(k,k) S(k) \right] \nonumber \\
  & & + \, a
  \left(q_\alpha q_\beta - q^2 \delta_{\alpha \beta}\right)
  \int  \frac{d^4 k}{(2 \pi)^4} {\rm Tr} \left[
  W_{2 \alpha}(k,k) S(k) \right]\Bigg\} + \mathcal{O}(a^2 q^4) 
  \nonumber 
\end{eqnarray} 
where $\Pi_{\alpha \beta}^{a b}(q; c_{SW}, 0)$ is the vacuum polarization tensor with no smearing, 
$W_1$ and $W_2$ are the Wilson quark gluon vertices defined in (\ref{eq:A3}), and the trace is now
only over the spin index. All $\omega^2$ terms have dropped out because they first appear at
$\mathcal{O}(a^2q^4)$. Calculating $\Pi_{\alpha \beta}^{a b}(q; c_{SW}, 0)$ in one loop
for $c_{SW}=1$ leads to the value of $d_f$  given in Eq.~(\ref{df}).

From power counting we would at first expect the integrals $\propto \,\omega$ in (\ref{Pilong}) 
to have values proportional to $1/a^2$ or $1/a^3$, and to make a finite contribution to $d_f$.
However we  show now that there is a perfect cancellation between the continuum-like diagram 
Fig.~\ref{fig4}(a) (the integrals involving $W_1$) and the tadpole contribution Fig.~\ref{fig4}(b) 
(those with $W_2$). To do this we use the identities
\begin{eqnarray} 
 \frac{ \partial}{\partial k_\mu} S(k) &=& - \, S(k) W_{1 \mu}(k,k) S(k)\,, \\
  \frac{ \partial}{\partial k_\mu} W_{1 \nu}(k,k) 
  &=& - \, a \,\delta_{\mu \nu} W_{2 \mu}(k,k)  
\end{eqnarray} 
which follow immediately from the definitions. Eq.~(\ref{Pilong}) becomes
\begin{eqnarray}
  \Pi_{\alpha \beta}^{a b}(q; c_{SW}, \omega) &=&
  \Pi_{\alpha \beta}^{a b}(q; c_{SW}, 0) + 2 \,\omega\, N_f\, \delta^{a b} g^2 a^2 \times\nonumber\\
  & & \bigg\{
  \sum_\mu\,\left(q_\alpha q_\mu - q^2 \delta_{\alpha \mu}\right)
  \int  \frac{d^4 k}{(2 \pi)^4}\,
  \frac{\partial}{\partial k_\beta} {\rm Tr} \left[
  W_{1 \mu}(k,k) S(k) \right] \label{Pishort} \\
  &&+\,
  \sum_\mu\,\left(q_\beta q_\mu - q^2 \delta_{\beta \mu}\right)
  \int  \frac{d^4 k}{(2 \pi)^4}\,
  \frac{\partial}{\partial k_\alpha} {\rm Tr} \left[
  W_{1 \mu}(k,k) S(k) \right]\bigg\}
  + \mathcal{O}(a^2 q^4) \nonumber \,.
\end{eqnarray}
The integrals are now zero because $W_1$ and $S$ are periodic, 
\begin{equation} 
  \int_{-\pi/a}^{\pi/a} d k_\alpha \frac{\partial}{\partial k_\alpha}
  {\rm Tr} \left[ W_{1 \mu}(k,k) S(k) \right]
  = {\rm Tr} \left[ W_{1 \mu}(k,k) S(k) \right]
  \Big|_{k_\alpha=-\pi/a}^{k_\alpha=\pi/a} = 0 \,. 
\end{equation} 
Thus we have proved that the vacuum polarization is independent of 
smearing the one-link part of the fermion action, 
\begin{equation} 
  \Pi_{\alpha \beta}^{a b}(q; c_{SW}, \omega) = 
  \Pi_{\alpha \beta}^{a b}(q; c_{SW}, 0)   + \mathcal{O}(a^2 q^4)
\end{equation}
which implies that $d_f$ depends on $r$ and $c_{SW}$, but not on $\omega$.

\end{document}